\documentclass[pdftex]{article}

\usepackage[a4paper]{geometry}

\usepackage[author-year]{amsrefs}
\usepackage{url}

\usepackage[T1]{fontenc}              
\usepackage[utf8]{inputenc}         
\usepackage[english]{babel}

\usepackage{amsthm,amsmath,amsfonts,amssymb} 
\usepackage{amsthm}                   
\usepackage{graphicx}                 
\usepackage{subfigure}                   
\usepackage{lscape}

\usepackage{listings}                 
\usepackage{tikz}

\usepackage{algorithm2e}
\usepackage{mathtools}
\usepackage{enumerate}

\usepackage[strict]{changepage}
\usepackage{relsize}
\usepackage{authblk}
\usepackage{epstopdf}
\usepackage{bbm}
          
\usepackage{standalone}

\def\eqd{\,{\buildrel d \over =}\,} 

\theoremstyle{plain}% default

\theoremstyle{definition}

\theoremstyle{remark}

\begin{document}
\title{Non-stationary Spatial Modelling with Applications to Spatial Prediction of Precipitation}
\author[1]{Geir-Arne Fuglstad\thanks{fuglstad@math.ntnu.no}}
\author[1]{Daniel Simpson}
\author[2]{Finn Lindgren}
\author[1]{Håvard Rue}
\affil[1]{Department of Mathematical Sciences, NTNU, Norway}
\affil[2]{Department of Mathematical Sciences, University of Bath, UK}
\date{June 3, 2013}
\maketitle

\begin{abstract}
A non-stationary spatial Gaussian random field (GRF) is described as
the solution of an inhomogeneous stochastic partial differential equation (SPDE),
where the covariance structure of the GRF is controlled by the coefficients in
the SPDE. This allows for a flexible way to vary the covariance structure,
where intuition about the resulting structure can be gained from the
local behaviour of the differential equation. Additionally, computations
can be done with computationally convenient Gaussian Markov random fields 
which approximate the true GRFs. The model is applied to a 
dataset of annual precipitation in the conterminous US. 
The non-stationary model performs better than a stationary model 
measured with both CRPS and the logarithmic scoring rule.

\end{abstract}

\section{Introduction}
\label{sec:Intro}
Classical geostatistical models are usually based on stationary Gaussian
random fields (GRFs) with covariates that capture important structure
in the mean. However, for environmental phenomena there is often no 
reason to believe that the covariance structures of the processes are
stationary. For example, the spatial 
distribution of precipitation is greatly affected by topography. Some of the 
effects can be captured in the mean, but other effects such as decreased dependence
between two locations because there is a mountain between them is something
which should enter in the covariance structure. The main focus of this paper
is on allowing a flexible model for the covariance structure of the process.
The goal is to improve spatial predictions at unmeasured locations.

\begin{figure}
	\centering
	\includegraphics[width=14cm]{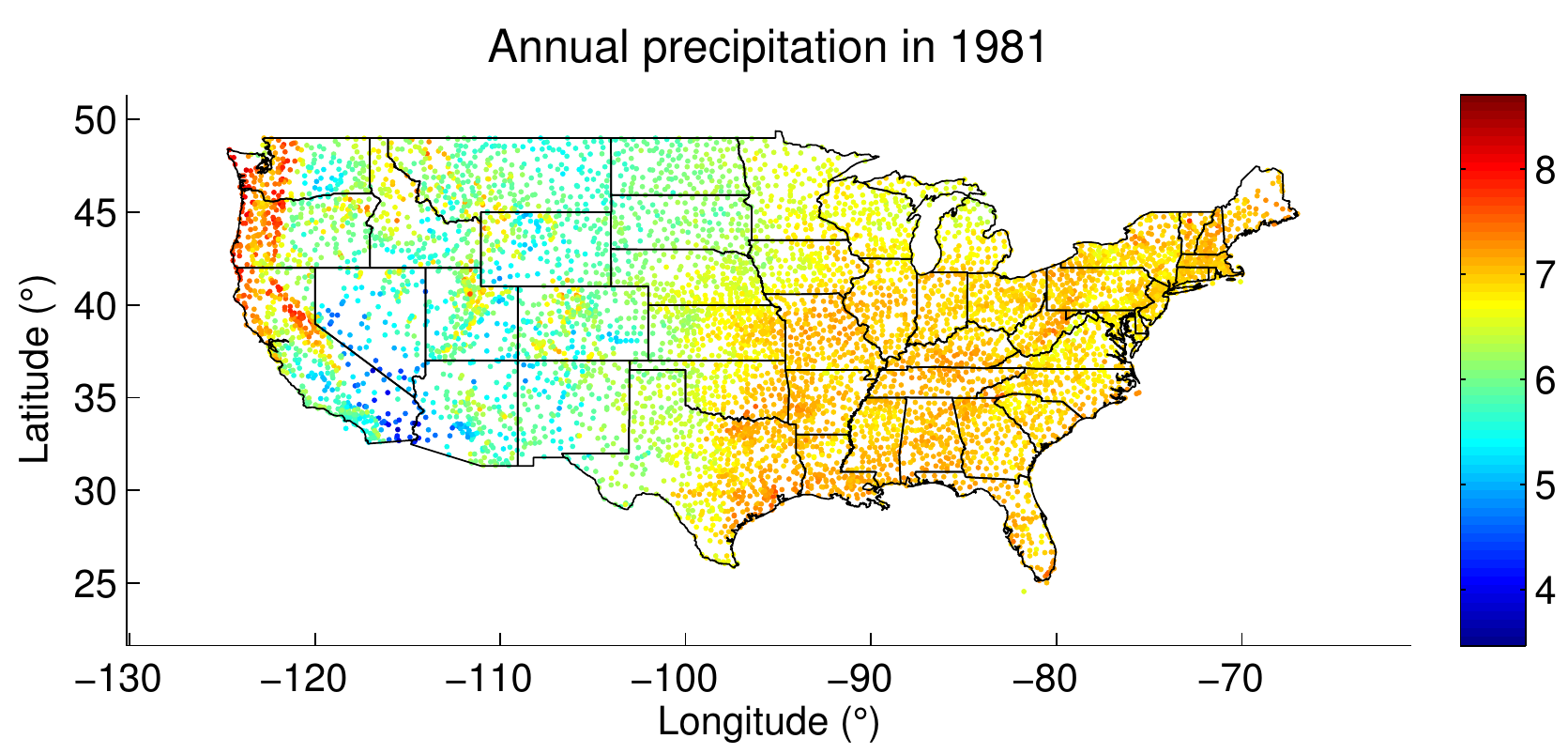}
	%\tikzsetnextfilename{Observations}
	%\input{TikzFigures/observation1981.tex}
	\caption{The logarithm of total yearly precipitation measured in millimetres at
			 7040 locations in the conterminous US for the year 1981.}
	\label{fig:observation}
\end{figure}

The dataset used consists of monthly precipitation measured in millimetres per 
month for the conterminous US for the years 1895--1997 and is available at the 
web page \url{http://www.image.ucar.edu/GSP/Data/US.monthly.met/}. 
There are a total of 11918 measurement stations, but measurements are only
available for some stations each month and the rest of the stations
have infill values~\cite{Johns2003}. The monthly data for 1981 is aggregated to yearly precipitation
for those stations which have measurements available for all months. 
This leaves a total of 7040 measurements of yearly precipitation in 1981.
After aggregating the data, the logarithm of each value is taken. This gives
the observations shown in Figure~\ref{fig:observation}. 
The only covariate available in the dataset is the elevation at each 
station, and since the focus
of the paper is on the covariance structure, no work was done to find new
covariates from other sources.
\ocite{Paciorek2006} previously used the same dataset to study a kernel-based
non-stationary model for annual precipitation, but their work was restricted
to the state of Colorado. 
%, by \ocite{Furrer2006} 
%to demonstrate their covariance tapering techniques and 
%by~\ocite{Daly2002} to create a gridded data product.

The traditional approaches to spatial modelling are based on covariance functions
and are severely limited by a cubic increase in computation time as a function
of the number of observations and prediction locations. In this paper we take a
different approach based on the connection between stochastic partial differential equations (SPDEs) 
and some classes of GRFs that was developed by~\ocite{Lindgren2011}.
The main computational benefit of this approach comes from a transition from a covariance 
function based representation to a Gaussian Markov random field (GMRF)~\cite{Rue2005} based formulation.
In a similar way as a spatial GMRF describes local behaviour for
a discretely indexed process, an SPDE describes local behaviour for a continuously indexed 
process. This continuous description of local behaviour can be transferred 
to a GMRF approximation
of the solution of the SPDE and gives a GMRF with a spatial Markovian structure that
can be exploited in computations.

\ocite{Lindgren2011} showed that Mat\'{e}rn-like GRFs can be constructed from a relatively simple
stationary SPDE, which can be thought of as a continuous linear filter driven by Gaussian white noise. From
this starting point it is possible to make a spatially varying linear filter that imposes different
smoothing of the Gaussian white noise at different locations. This leads to a non-stationary GRF whose
global covariance structure is modelled through the local behaviour at each location. Such SPDE-based
non-stationary GRFs have previously been applied to global ozone data~\cite{Bolin2011} 
and annual precipitation data for Norway~\cite{Rikke2013}. Their models preserved the computational 
benefits of GMRFs, but allowed for non-stationary covariance structures. 
The model used in this paper is constructed in a similar fashion, but with a focus on varying 
the local anisotropy as in~\ocite{Fuglstad2013}.

The closest among the more well-known methods for non-stationary spatial modelling
is the deformation method~\cite{Sampson1992}, where an isotropic process is made
non-stationary by deforming the base space. However, in the method presented in this
paper it is not the deformation itself that is modelled, but rather the distance measure
in the space. Loosely speaking, one controls local range and anisotropy. This
usually leads to distance measures that cannot be described by a deformation of
\(\mathbb{R}^2\) and require embeddings into higher dimensional spaces. But all deformations
can be described by a change of the distance measure. The original formulation of the deformation
method has later been extended to Bayesian 
variants~\cite{Damian2001, Damian2003,Schmidt2003, Schmidt2011}. 

Another type of methods with some connection to the SPDE-based models 
is the kernel methods~\cite{Higdon1998, Paciorek2006} 
in which a spatially varying kernel is convolved with Gaussian white noise. The
solutions of the SPDE can be formulated in the same manner, but this is
not practical and  would lead to something more directly connected to the covariance matrix as opposed to the 
conditional formulation where the precision matrix is directly available.
The already mentioned methods are the most closely relevant ones, but there is a large
literature on other non-stationary 
methods~\cite{Haas1990b, Haas1990a, Kim2005, Fuentes2001, Fuentes2002b, Fuentes2002a }.

The paper is divided into five sections. Section~\ref{sec:Theory} describes the
theoretical basis for the spatial model. The connection between the coefficients
that control the SPDE and the resulting GRF is discussed, and a way to parametrize
the GRF is presented. In Section~\ref{sec:FullModel}, a Bayesian hierarchical model using
the GRF is constructed and the resulting posterior distribution is 
explicitly given. Then in Section~\ref{sec:Application} the hierarchical model
is applied to annual precipitation in the conterminous US for 1981.
The predictions of the non-stationary model is compared to the stationary model.
Lastly, the paper ends with discussion and concluding remarks in 
Section~\ref{sec:Discussion}.

\section{The spatial model}
\label{sec:Theory}

\subsection{The SPDE-based model}
\label{sec:GRF}
\begin{figure}
	\centering
	\includegraphics[width=10cm]{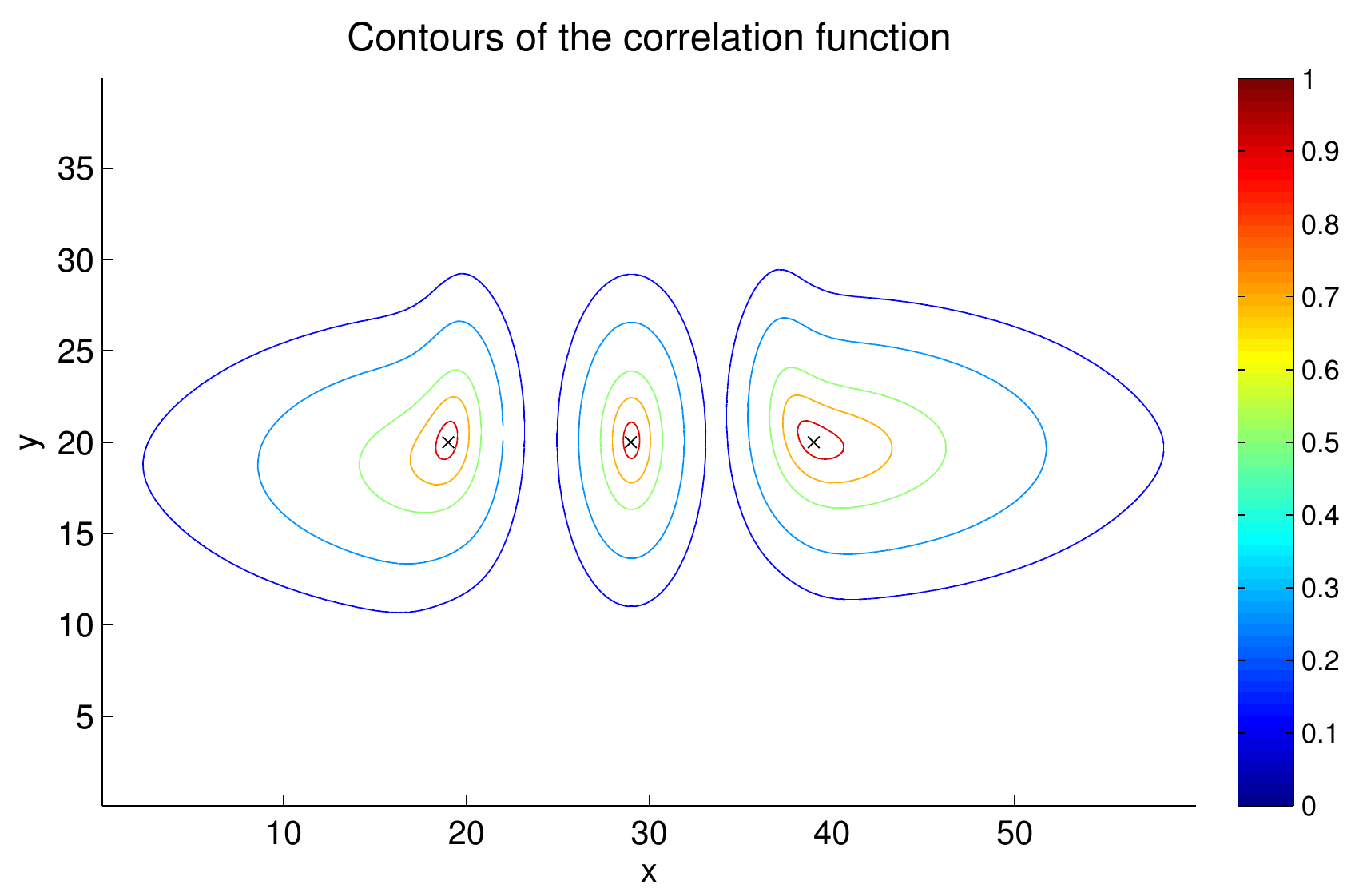}
	\caption{Example of a correlation function caused by varying local behaviour.
			 For each location marked with a black cross, the 0.9, 0.7, 0.5, 0.25 
			 and 0.12 level contours of the correlation function are shown.}
	\label{fig:CorrExample}
\end{figure}

A good spatial model should provide a useful way to model spatial phenomena. 
For a non-stationary model,
it is not easy to create a global covariance function when one only has 
intuition about local behaviour. Consider the situation in Figure~\ref{fig:CorrExample}.
The left hand side and right hand side has locally large ``range'' in the horizontal
direction and somewhat shorter ``range'' in the vertical direction, and the middle area has
locally much shorter ``range'' in the horizontal direction, but slightly longer in the
vertical direction. From the figure one can see that for the point in the middle,
the chosen contours look more or less unaffected by the two other regions since 
they are fully contained in the middle region, but that for the point on the left hand
side and the point on the right hand side, there is much skewness introduced by the
transition into a different region. 

It would be very hard to directly construct a fitting correlation
function for this situation. Therefore, this example provides a strong case for the 
use of SPDE-based models. This is exactly the type of behaviour they can describe. 
With the SPDE-based approach one does not try to model the global behaviour
directly, but rather how the the ``range'' behaves locally. This
implies that specifying a specific correlation between two locations
is hard, but this is besides the point. The correlation should be determined 
by what happens between the locations. That is whether there are plains
or lakes where one might believe that the correlation decreases slowly or a mountain 
where one might believe that the correlation decreases quickly. A great thing about this 
type of local specification is that it naturally leads to a spatial GMRF that 
has good computational properties. Since each location is conditionally
dependent only on locations close to itself, the precision matrix is sparse.

The starting point for the non-stationary SPDE-based model 
is the stationary SPDE introduced in~\ocite{Lindgren2011},
\begin{equation}
	(\kappa^2-\nabla\cdot\nabla)u(\boldsymbol{s}) = \sigma \mathcal{W}(\boldsymbol{s}), \qquad \boldsymbol{s}\in\mathbb{R}^2,
	\label{eq:LindgrenSPDE}
\end{equation}
where \(\kappa>0\) and \(\sigma>0\) are constants, 
\(\nabla = (\frac{\partial}{\partial x}, \frac{\partial}{\partial y})^\mathrm{T}\) and \(\mathcal{W}\) 
is standard Gaussian white noise. This SPDE basically describes the GRF \(u\) as a 
smoothed version of the Gaussian white noise on the right hand side. \ocite{Whittle1954} showed
that any stationary solution of this SPDE has the Mat\'{e}rn covariance function
\begin{equation}
	r(\boldsymbol{s}_1, \boldsymbol{s}_2) = \frac{\sigma^2}{4\pi \kappa^2} (\kappa \lvert\lvert \boldsymbol{s}_2-\boldsymbol{s}_1 \rvert\rvert)K_{1}(\kappa \lvert\lvert \boldsymbol{s}_2-\boldsymbol{s}_1 \rvert\rvert),
	\label{eq:MaternCov}
\end{equation} 
where \(K_1\) is the modified Bessel function of second kind, order 1.
This covariance function is a member of
one of the more commonly used families of covariance functions, and one can
see from Equation~\eqref{eq:MaternCov} that one can first use \(\kappa\) 
to select the range and then \(\sigma\) to achieve the desired marginal
variance.

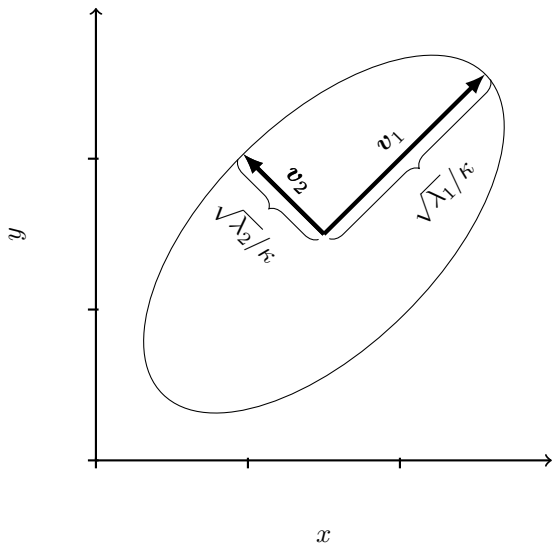
\begin{figure}
	\centering
	\begin{tikzpicture}
	\usetikzlibrary{decorations.pathreplacing}
	% Axis
	\draw[->, thick] (0,0) -- coordinate (x axis mid) (6cm,0);
    \draw[->, thick] (0,0) -- coordinate (y axis mid) (0,6cm);
    
    % Ticks on axes
    \draw[thick] (0, 1pt) -- (0, -3pt);
    \draw[thick] (2, 1pt) -- (2, -3pt);
    \draw[thick] (4, 1pt) -- (4, -3pt);
    \draw[thick] (1pt, 0) -- (-3pt, 0);
    \draw[thick] (1pt, 2) -- (-3pt, 2);
    \draw[thick] (1pt, 4) -- (-3pt, 4);

    % Labels
    \node[below=0.8cm] at (x axis mid) {$x$};
    \node[rotate=90, above = 0.8cm] at (y axis mid) {$y$};

    % Vectors
    \draw [ultra thick, -latex] (3cm, 3cm) -- (5.121, 5.121) node [midway, above, sloped] (TextNode) {$\boldsymbol{v}_1$};
    \draw [decorate, decoration={brace, amplitude=5pt}, yshift = -2pt, xshift = +2pt] (5.121, 5.121) -- (3,3) node [midway, below, sloped, yshift=-8pt] {$\sqrt{\lambda_1}/\kappa$};
    \draw [ultra thick, -latex] (3cm, 3cm) -- (1.939, 4.061) node [midway, above, sloped] (TextNode2) {$\boldsymbol{v}_2$};
	\draw[rotate around={-45:(3,3)}] (3,3) ellipse (1.5cm and 3cm);
	\draw [decorate, decoration={brace, amplitude=5pt}, yshift = -2pt, xshift = -2pt] (3,3) -- (1.939,4.061) node [midway, below, sloped, yshift=-8pt] {$\sqrt{\lambda_2}/\kappa$};
    \draw [ultra thick, -latex] (3cm, 3cm) -- (1.939, 4.061) node [midway, above, sloped] (TextNode2) {$\boldsymbol{v}_2$};
\end{tikzpicture}
	\caption{Isocorrelation curve for the 0.6 level, where \((\lambda_1, \boldsymbol{v}_1)\) and \((\lambda_2, \boldsymbol{v}_2)\) are the eigenpairs of \(\mathbf{H}\).}
	\label{fig:CovFunction}
\end{figure}

The next step is to generate a GRF with an anisotropic Mat\'{e}rn covariance function. The
reason behind the isotropy in the SPDE above is that the Laplacian, 
\(\Delta = \nabla\cdot\nabla\) is invariant to a change of coordinates that 
only involves rotation and translation. To change this a \(2\times 2\)
matrix \(\mathbf{H}>0\) is introduced into the operator to give the SPDE
\begin{equation}
	(\kappa^2-\nabla\cdot\mathbf{H}\nabla)u(\boldsymbol{s}) = \sigma \mathcal{W}(\boldsymbol{s}).
	\label{eq:AniSPDE}
\end{equation}
This choice is closely related to a change of coordinates 
\(\tilde{\boldsymbol{s}}=\mathbf{H}^{1/2}\boldsymbol{s}\) and gives the covariance function
\begin{equation}
	r(\boldsymbol{s}_1, \boldsymbol{s}_2) = \frac{\sigma^2}{4\pi\kappa^2\sqrt{\det(\mathbf{H})}} (\kappa \lvert\lvert \mathbf{H}^{-1/2}(\boldsymbol{s}_2-\boldsymbol{s}_1) \rvert\rvert)K_{1}(\kappa \lvert\lvert \mathbf{H}^{-1/2}(\boldsymbol{s}_2-\boldsymbol{s}_1) \rvert\rvert).
	\label{eq:aniMatern}
\end{equation}
Compared to Equation~\eqref{eq:MaternCov} there is a change in the marginal variance and
a directionality is introduced through a distance measure that is different than the 
standard Euclidean distance. Figure~\ref{fig:CovFunction} shows how the eigenpairs of \(\mathbf{H}\)
and the value of \(\kappa\) act together to control range. One can see that this leads to
elliptic isocovariance curves.
In what follows \(\sigma\) is assumed to be equal to 1 since the marginal variance can
be controlled by varying \(\kappa^2\) and \(\mathbf{H}\) together.

The final step is to construct a non-stationary GRF where the local behaviour at each location
is governed by SPDE~\eqref{eq:AniSPDE}, but \(\sigma = 1\) and the values of \(\kappa^2\) and
\(\mathbf{H}\) vary over the domain. The intention is to create a GRF by 
chaining together processes with different covariance structures. The SPDE becomes
\begin{equation}
	(\kappa^2(\boldsymbol{s})-\nabla\cdot\mathbf{H}(\boldsymbol{s})\nabla)u(\boldsymbol{s}) = \mathcal{W}(\boldsymbol{s}).
	\label{eq:FinalSPDE}
\end{equation}
For technical reasons concerned with the discretization in the next section, \(\kappa^2\) 
is required to be continuous and \(\mathbf{H}\)
is required to be continuously differentiable. See~\ocite{Fuglstad2013} for a study 
of the case where \(\kappa^2\) is constant and \(\mathbf{H}\) varies.

\subsection{Gaussian Markov random field approximation}
SPDE~\eqref{eq:FinalSPDE} describes the covariance
structure of a GRF, but the information must be brought into a form which is
useful for computations. The first thing to notice is that the operator 
in front of \(u\) only contains multiplications with functions and
first order and second order derivatives. All of these operations involve only
the local properties of \(u\) at each location. This means that if \(u\) is 
discretized, the corresponding discretized operators (matrices) should 
only involve variables close to each other. This leads to a sparse GMRF 
which possesses approximately the same covariance structure as \(u\).

The first step in creating the GMRF is to restrict 
SPDE~\eqref{eq:FinalSPDE} to a bounded domain,
\[
	(\kappa^2(\boldsymbol{s})-\nabla\cdot\mathbf{H}(\boldsymbol{s})\nabla)u(\boldsymbol{s}) = \mathcal{W}(\boldsymbol{s}), \qquad \boldsymbol{s}\in\mathcal{D} = [A_1,B_1]\times[A_2, B_2]\subset\mathbb{R}^2,
\]
where \(B_1 > A_1\) and \(B_2 > A_2\). This restriction 
necessitates a boundary condition to make the distribution useful and 
proper. For technical reasons the boundary condition chosen is 
zero flux across the boundaries. The derivation of a discretized version
of this SPDE on a grid is somewhat involved, but for periodic boundary 
conditions the derivation can be found in~\ocite{Fuglstad2013}. The boundary 
conditions in this problem involve a slight change in that derivation. 

For a regular \(m \times n\) grid of \(\mathcal{D}\), the end result is the matrix
equation
\[
	\mathbf{A}(\kappa^2, \mathbf{H}) \boldsymbol{u} = \frac{1}{\sqrt{V}}\boldsymbol{z},
\]
where \(V\) is the area of each cell in the grid, \(\boldsymbol{u}\)
corresponds to the values of u on the cells in the regular grid stacked column-wise, 
\(\boldsymbol{z}\sim\mathcal{N}_{mn}(\boldsymbol{0}, \mathbf{I}_{mn})\) and
\(\mathbf{A}(\kappa^2, \mathbf{H})\) is a discretized version of 
\((\kappa^2-\nabla\cdot\mathbf{H}\nabla)\). This matrix equation leads to the distribution
\begin{equation}
	\boldsymbol{u} \sim \mathcal{N}_{mn}(\boldsymbol{0}, \mathbf{Q}(\kappa^2, \mathbf{H})^{-1}),
	\label{eq:GMRFeq}
\end{equation}
where \(\mathbf{Q}(\kappa^2, \mathbf{H}) = \mathbf{A}(\kappa^2, \mathbf{H})^\mathrm{T}\mathbf{A}(\kappa^2, \mathbf{H})V\). The precision matrix has up to 25 non-zero elements in each row,
corresponding to the point itself, its eight closest neighbours and the eight 
closest neighbours of the eight closest neighbours.

This type of construction alleviates one of the largest problems 
with GMRFs, namely that they are hard to specify.
The computational benefits of spatial GMRFs are well known, but a GMRF needs to
be constructed through its conditional distributions. This is notoriously hard
to do for non-stationary models. But with the type of derivation outlined above
it is possible to model the problem with an SPDE and then automatically find
a GMRF corresponding to this model, but with the computational benefits of
GMRFs preserved.

\subsection{Decomposition of \(\mathbf{H}\)}
The function \(\mathbf{H}\) must give positive definite \(2 \times 2\) matrices
at each location. One usual way to do this is to use two strictly positive functions
\(\lambda_1\) and \(\lambda_2\) for the eigenvalues and a function \(\phi\) for the
angle between the \(x\)-axis and the eigenvector associated with \(\lambda_1\).
However, with a slight re-parametrization
\(\mathbf{H}\) can be written as the sum of an isotropic effect, described by a constant
times the identity matrix, plus an additional anisotropic effect, described by
direction and magnitude. Write \(\mathbf{H}\) as a function of
the  scalar functions \(\gamma\), \(v_x\) and \(v_y\) by
\[
	\mathbf{H}(\boldsymbol{s}) = \gamma(\boldsymbol{s})\mathbf{I}_2 + \begin{bmatrix} v_x(\boldsymbol{s}) \\ v_y(\boldsymbol{s}) \end{bmatrix} \begin{bmatrix} v_x(\boldsymbol{s}) & v_y(\boldsymbol{s}) \end{bmatrix},
\]
where \(\gamma\) is required to be strictly positive. 
The eigendecomposition of this matrix has eigenvalue
\(\lambda_1(\boldsymbol{s}) = \gamma(\boldsymbol{s})\) with eigenvector
\(\boldsymbol{v}_1(\boldsymbol{s}) = (v_x(\boldsymbol{s}), v_y(\boldsymbol{s}))\)
and eigenvalue \(\lambda_2(\boldsymbol{s}) = \gamma(\boldsymbol{s})+v_x(\boldsymbol{s})^2+v_y(\boldsymbol{s})^2\) with eigenvector \(\boldsymbol{v}_2(\boldsymbol{s}) =(-v_y(\boldsymbol{s}), v_x(\boldsymbol{s}))\). From Figure~\ref{fig:CovFunction} this means that for
a stationary model, \(\gamma\) affects the length of the shortest semi-axis 
of the isocorrelation curves and \(\boldsymbol{v}\) specifies the direction of
and how much larger the longest semi-axis is.

\subsection{Parametrization of the model}
Since the focus lies on allowing flexible covariance structures, some
representation of the functions \(\kappa^2\), \(\gamma\), \(v_x\) and \(v_y\) is
needed. To ensure positivity of \(\kappa^2\) and \(\gamma\), they are first 
transformed into \(\log(\kappa^2)\) and \(\log(\gamma)\). 
The choice was made to make \(\log(\kappa^2)\), \(\log(\gamma)\), \(v_x\) and \(v_y\)
Gaussian processes a priori. This requires both a finite dimensional representation of 
each function and appropriate priors to connect the parameters in each function to each other.  
The steps that follow are the same for each function. Therefore, they are
only shown for \(\log(\kappa^2)\).

A priori \(\log(\kappa^2)\) is given the distribution generated from the SPDE
\begin{equation}
	-\Delta \log(\kappa^2(\boldsymbol{s})) = \mathcal{W}_\kappa(\boldsymbol{s})/\sqrt{\tau_\kappa}, \qquad \boldsymbol{s}\in\mathcal{D},
	\label{eq:Prior}
\end{equation}
where \(\tau_\kappa > 0\) is a scale hyperparameter, with an additional requirement
of zero derivatives at the edges. This extra requirement is used to restrict
the resulting distribution so it is only invariant to the addition of
a constant function, and the hyperparameter
is used to regulate how much \(\log(\kappa^2)\) can vary from a constant function.
The prior defined through SPDE~\eqref{eq:Prior} is in this paper called a 
two-dimensional second-order random walk due to its similarity 
to a one-dimensional second-order random walk~\cite{Lindgren2008}.

The next step is to expand \(\log(\kappa^2)\) in a basis through a linear
combination of basis functions,
\[
	\log(\kappa^2(\boldsymbol{s})) = \sum_{i=1}^{k}\sum_{j=1}^{l} \alpha_{ij} f_{ij}(\boldsymbol{s}),
\]
where \(\{\alpha_{ij}\}\) are the parameters and
\(\{f_{ij}\}\) are real-valued basis functions. For convenience, the basis
is chosen in such a way that all basis functions satisfy the boundary conditions
specified in SPDE~\eqref{eq:Prior}. If this is done, one does not have
to think more about the boundary condition. The remaining tasks are then
to decide which basis functions to use and what distribution the parameters should be given.

Due to a desire to make
\(\mathbf{H}\) continuously differentiable and a desire to have ``local'' basis
functions, the basis functions are chosen to be based on 2-dimensional, 
second-order B-splines (piecewise-quadratic functions). The basis is constructed
as a tensor product of two 1-dimensional B-spline bases constrained to 
satifsy the boundary condition.

The final step is to determine a Gaussian distribution for the parameters
such that the distribution of \(\log(\kappa^2)\) is close to a solution of
SPDE~\eqref{eq:Prior}. The approach taken is based on a least-squares 
formulation of the solution
and is described in Appendix~\ref{app:Prior}. Let \(\boldsymbol{\alpha}\) 
be the \(\{\alpha_{ij}\}\) parameters stacked row-wise, then the
result is that \(\boldsymbol{\alpha}\) should be given
a zero-mean Gaussian distribution with precision matrix 
\(\tau_\kappa\mathbf{Q}_{\mathrm{RW2}}\). This matrix has rank \((kl-1)\) and
the distribution is invariant to the addition of a vector of only the same values, 
but for convenience the 
distribution will still be written as \(\boldsymbol{\alpha}\sim\mathcal{N}_{kl}(\boldsymbol{0}, \mathbf{Q}_{\mathrm{RW2}}^{-1}/\tau_\kappa)\).

\section{Hierarchical model}
\label{sec:FullModel}

\subsection{Full model}
\label{sec:fullModel}
Observations \(y_1, y_2, \ldots, y_N\) are made at locations 
\(\boldsymbol{s}_1, \boldsymbol{s}_2, \ldots, \boldsymbol{s}_N\). The observed
value at each location is assumed to be the sum of a fixed
effect due to covariates, a spatial ``smooth'' effect and a random effect. The
covariates at location \(\boldsymbol{s}_i\) is described by the \(p\)-dimensional
row vector \(\boldsymbol{x}(\boldsymbol{s}_i)^\mathrm{T}\) and the spatial field is
denoted by \(u\). This gives the equation
\[
	y_i = \boldsymbol{x}(\boldsymbol{s}_i)^\mathrm{T}\boldsymbol{\beta} + u(\boldsymbol{s}_i) + \epsilon_i,
\]
where \(\boldsymbol{\beta}\) is a \(p\)-variate random vector for the 
coefficients of the covariates and 
\(\epsilon_i \sim \mathcal{N}(0, 1/\tau_{\mathrm{noise}})\) is the random
effect for observation \(i\). 

The \(u\) is modelled as in Section~\ref{sec:Theory} and a
GMRF approximation is introduced for computations. In this GMRF 
approximation the domain is divided into a regular grid consisting
of rectangular cells and each element of the GMRF approximation
describes the average value on one of these cells. So
\(u(\boldsymbol{s}_i)\) is replaced with the approximation
\(\boldsymbol{e}(\boldsymbol{s}_i)^\mathrm{T}\boldsymbol{u}\), where 
\(\boldsymbol{e}(\boldsymbol{s}_i)^\mathrm{T}\) is the \(mn\)-dimensional
row vector selecting the element of \(\boldsymbol{u}\) which corresponds
to the cell which contains location \(\boldsymbol{s}_i\). In total, this gives
\begin{equation}
	\boldsymbol{y} = \mathbf{X}\boldsymbol{\beta}+\mathbf{E}\boldsymbol{u} + \boldsymbol{\epsilon},
	\label{eq:DataEquation}
\end{equation}
where \(\boldsymbol{y} = (y_1, y_2, \ldots, y_N)\), the matrix \(\mathbf{X}\) 
has \(\boldsymbol{x}(\boldsymbol{s}_1)^\mathrm{T},\ldots,\boldsymbol{x}(\boldsymbol{s}_N)^\mathrm{T}\)
as rows and the matrix \(\mathbf{E}\) has 
\(\boldsymbol{e}(\boldsymbol{s}_1)^\mathrm{T},\ldots,\boldsymbol{e}(\boldsymbol{s}_N)^\mathrm{T}\) as 
rows. The
model for the observations can also be written in the form
\[
	\boldsymbol{y}|\boldsymbol{\beta},\boldsymbol{u},\log(\tau_\mathrm{noise}) \sim \mathcal{N}_N(\mathbf{X}\boldsymbol{\beta} + \mathbf{E}\boldsymbol{u}, \mathbf{I}_N/\tau_\mathrm{noise} ).
\]
The parameter \(\tau_{\mathrm{noise}}\) 
acts as the precision of a joint effect 
from measurement noise and small scale spatial variation~\cite{Diggle2007}.

This can be turned into a full Bayesian model by providing priors for the two remaining
parameters. The \(p\)-dimensional random variable \(\boldsymbol{\beta}\) is given a weak
Gaussian prior,
\[
	\boldsymbol{\beta} \sim \mathrm{N}_p(\boldsymbol{0}, \mathbf{I}_p/\tau_{\beta}),
\]
and the precision parameter \(\tau_{\mathrm{noise}}\) is given an improper, uniform prior on log-scale,
\[
	\log(\tau_{\mathrm{noise}}) \sim \mathcal{U}(0,\infty).
\]

To describe the full hierarchical model it is necessary to introduce some symbols to denote the
parameters and hyperparameters in the spatial field \(u\). Denote the parameters that
control \(\log(\kappa^2)\), \(\log(\gamma)\), \(v_x\) and \(v_y\) by 
\(\boldsymbol{\alpha}_1\), \(\boldsymbol{\alpha}_2\), \(\boldsymbol{\alpha}_3\) and
\(\boldsymbol{\alpha}_4\), respectively. Further, denote the corresponding scale hyperparameters that
controls the degree of smoothing for each function by \(\tau_1\), \(\tau_2\), \(\tau_3\) and \(\tau_4\).
With this notation the full model becomes
\begin{align*}
	\text{Stage 1: } & \boldsymbol{y} | \boldsymbol{\beta}, \boldsymbol{u}, \log(\tau_{\mathrm{noise}}) \sim \mathcal{N}_N(\mathbf{X}\boldsymbol{\beta}+\mathbf{E}\boldsymbol{u}, \mathbf{I}_N/\tau_{\mathrm{noise}}) \\
	\text{Stage 2: } &  \boldsymbol{u}|\boldsymbol{\alpha}_1, \boldsymbol{\alpha}_2, \boldsymbol{\alpha}_3, \boldsymbol{\alpha}_4 \sim \mathcal{N}_{nm}(\boldsymbol{0}, \mathbf{Q}^{-1}), \, \, \, \, \boldsymbol{\beta} \sim \mathcal{N}_p(\boldsymbol{0}, \mathbf{I}_p/\tau_{\beta})\\
	\text{Stage 3: } &\log(\tau_{\mathrm{noise}}) \sim \mathcal{U}(0, \infty), \,\,\,\, \boldsymbol{\alpha}_i| \tau_i \sim \mathcal{N}_{kl}(\boldsymbol{0}, \mathbf{Q}_{\mathrm{RW2}}^{-1}/\tau_i) \,\, \text{for}\,\, i = 1,2,3,4,
\end{align*}
%% TODO
% Need to do something about the specification of \alpha_i | \tau_i, 
% Q is not invertible so it is inconvinient to write in this form
where \(\tau_1\), \(\tau_2\), \(\tau_3\), \(\tau_4\) and \(\tau_\beta\) are hyperparameters 
that must be pre-selected in some way.  

\subsection{Posterior distribution and inference}
Denote the covariance parameters by 
\(\boldsymbol{\theta} = (\boldsymbol{\alpha}_1, \boldsymbol{\alpha}_2, \boldsymbol{\alpha}_3, \boldsymbol{\alpha}_4, \log(\tau_{\mathrm{noise}}))\),  
then the posterior distribution of 
interest is \(\pi(\boldsymbol{\theta}|\boldsymbol{y})\). The derivation of this
distribution involves integrating out Stage~2 of the hierarchical model. 
It is possible to do this explicitly (up to a constant) due to the fact that all 
distributions are Gaussian given the covariance parameters. 

First, treat the fixed effect and the spatial effect together with
\(\boldsymbol{z} = (\boldsymbol{u}^\mathrm{T}, \boldsymbol{\beta}^\mathrm{T})\). This leads to
\[
	\boldsymbol{z}|\boldsymbol{\theta}  \sim \mathcal{N}_{mn+p}(\boldsymbol{0}, \mathbf{Q}_{z}^{-1})
\]
and
\[
	\boldsymbol{y} | \boldsymbol{z}, \boldsymbol{\theta} \sim \mathcal{N}_{N}(\mathbf{S}\boldsymbol{z}, \mathbf{I}_N/\tau_{\mathrm{noise}}),
\]
where
\[
	\mathbf{S} = \begin{bmatrix} \mathbf{E} & \mathbf{X} \end{bmatrix}
\]
and
\[
	\mathbf{Q}_z = \begin{bmatrix} \mathbf{Q} & \mathbf{0} \\ \mathbf{0} & \tau_\beta \mathbf{I}_p \end{bmatrix}.
\]
Then apply the Bayesian formula to integrate out \(\boldsymbol{z}\) from the 
joint distribution of \(\boldsymbol{y}\), \(\boldsymbol{z}\) and 
\(\boldsymbol{\theta}\) as shown in Appendix~\ref{app:CondDist}. This gives
the full log-posterior
\begin{align}
\notag \log(\pi(\boldsymbol{\theta}|\boldsymbol{y})) &= \mathrm{Const} -\frac{1}{2}\sum_{i=1}^4\boldsymbol{\alpha}_i^\mathrm{T}\mathbf{Q}_{\mathrm{RW2}}\boldsymbol{\alpha}_i\cdot\tau_i +\frac{1}{2}\log(\det(\mathbf{Q}_z)) +\frac{N}{2}\log(\tau_\mathrm{noise})+ \\
	&-\frac{1}{2}\log(\det(\mathbf{Q}_\mathrm{C}))-\frac{1}{2}\boldsymbol{\mu}_\mathrm{C}^\mathrm{T}\mathbf{Q}_z\boldsymbol{\mu}_\mathrm{C}-\frac{\tau_\mathrm{noise}}{2}(\boldsymbol{y}-\mathbf{S}\boldsymbol{\mu}_\mathrm{C})^\mathrm{T}(\boldsymbol{y}-\mathbf{S}\boldsymbol{\mu}_\mathrm{C}),
	\label{eq:fullPost}
\end{align}
where \(\mathbf{Q}_\mathrm{C} = \mathbf{Q}_z+\mathbf{S}^\mathrm{T}\mathbf{S}\cdot\tau_\mathrm{noise}\)
and \(\boldsymbol{\mu}_\mathrm{C} = \mathbf{Q}_\mathrm{C}^{-1}\mathbf{S}^\mathrm{T}\boldsymbol{y}\cdot\tau_\mathrm{noise}\).

The properties of \(\boldsymbol{\theta}|\boldsymbol{y}\) are not easily
available from Equation~\eqref{eq:fullPost}. As such, the inference and the prediction
are done in something close to an empirical Bayes setting. 
The parameters are first estimated by the maximum a posteriori (MAP) estimator,
\({\skew{3}\hat{\boldsymbol{\theta}}}\), and then the values at new locations
\(\boldsymbol{y}^*\) are predicted with 
\(\boldsymbol{y}^*|{\skew{3}\hat{\boldsymbol{\theta}}}, \boldsymbol{y}\). However, the 
hyperparameters that control the priors for the covariance parameters are 
very hard to estimate. There is not enough information to estimate
the hyperparameters in the third stage of the hierarchical model. 
Thus they are selected with a cross-validation
procedure based on a score for the predictions. 

During implementation it 
became apparent that an analytic expression for the gradient was needed for 
the optimization to converge. Its form is given in Appendix~\ref{app:Gradient},
and its value can be computed for less cost than a finite 
difference approximation of the gradient with the number of parameters used in 
the application in this paper. 
The calculations require the use of techniques for calculating only parts of the 
inverse of a sparse precision matrix~\cite{Gelfand2010}*{Sections~12.1.7.10--12.1.7.12}.

\section{Application}
\label{sec:Application}
In this section the models and the estimation procedures presented in 
Sections~\ref{sec:Theory} and~\ref{sec:FullModel} are applied to
annual precipitation data for the conterminous US. A stationary and
a non-stationary model is fitted to the data and the quality of the
predictions are compared.

\subsection{The dataset}
The dataset is described in the introduction and consists of log-transformed
values of the annual precipitation measured in millimetres in 1981 for the
conterminous US for 7040 measurement stations. The values are shown in
Figure~\ref{fig:observation}. The elevation of each station is available and
is used together with a joint mean for the fixed effect. This means that the
design matrix, \(\mathbf{X}\), in Equation~\eqref{eq:DataEquation}
contains two columns. The first column contains
only ones, and corresponds to the joint mean, and the second column contains elevations
measured in kilometres. The coefficients of the fixed effect, \(\boldsymbol{\beta}\),
is given the vague prior
\(\boldsymbol{\beta} \sim \mathcal{N}_2(\boldsymbol{0}, \mathbf{I}_2\cdot 10^{4})\).

\subsection{Stationary model}
\label{sec:StatModel}
The spatial effect is constructed on a rectangular domain with 
longitudes from \(130.15\,^\circ\mathrm{W}\) to \(60.85\,^\circ\mathrm{W}\)
and latitudes from \(21.65\,^\circ \mathrm{N}\) to \(51.35\,^\circ \mathrm{N}\). 
This is larger that the actual size of the conterminous US as can be seen
in Figure~\ref{fig:observation}. This is done to reduce
boundary effects. The domain is discrectized into a \(400 \times 200\) grid
and the parameters \(\log(\kappa^2)\), \(\log(\gamma)\), \(v_x\), \(v_y\) and
\(\log(\tau_{\mathrm{noise}})\) are estimated. In this case the second order random walk
prior is not used as no basis (except a constant) is needed for the functions.
Each parameter is given a uniform prior and the parameters are estimated with
MAP estimates. The estimated values with associated approximate standard deviations
are shown in Table~\ref{tab:StatPar}. The approximate standard deviations 
are calculated from the observed information matrix.

\begin{table}
	\centering
	\caption{\sf Estimated values of the parameters and associated approximate standard deviations
			 for the stationary model.}
	\begin{tabular}{lll}
		\textbf{Parameter}			& \textbf{Estimate}	& \textbf{Standard deviation} \\
		\(\log(\kappa^2)\)			& \(-1.75\)			& \(0.15\)	\\
		\(\log(\gamma)\)			& \(-0.272\)		& \(0.042\)	\\
		\(v_x\)						& \(0.477\)			& \(0.053\) \\
		\(v_y\)						& \(-0.313\)		& \(0.057\)	\\
		\(\log(\tau_{\mathrm{noise}})\)	& \(4.266\)			& \(0.030\)
	\end{tabular}
	\label{tab:StatPar}
\end{table}

\begin{figure}
	\centering
	\includegraphics[width=10cm]{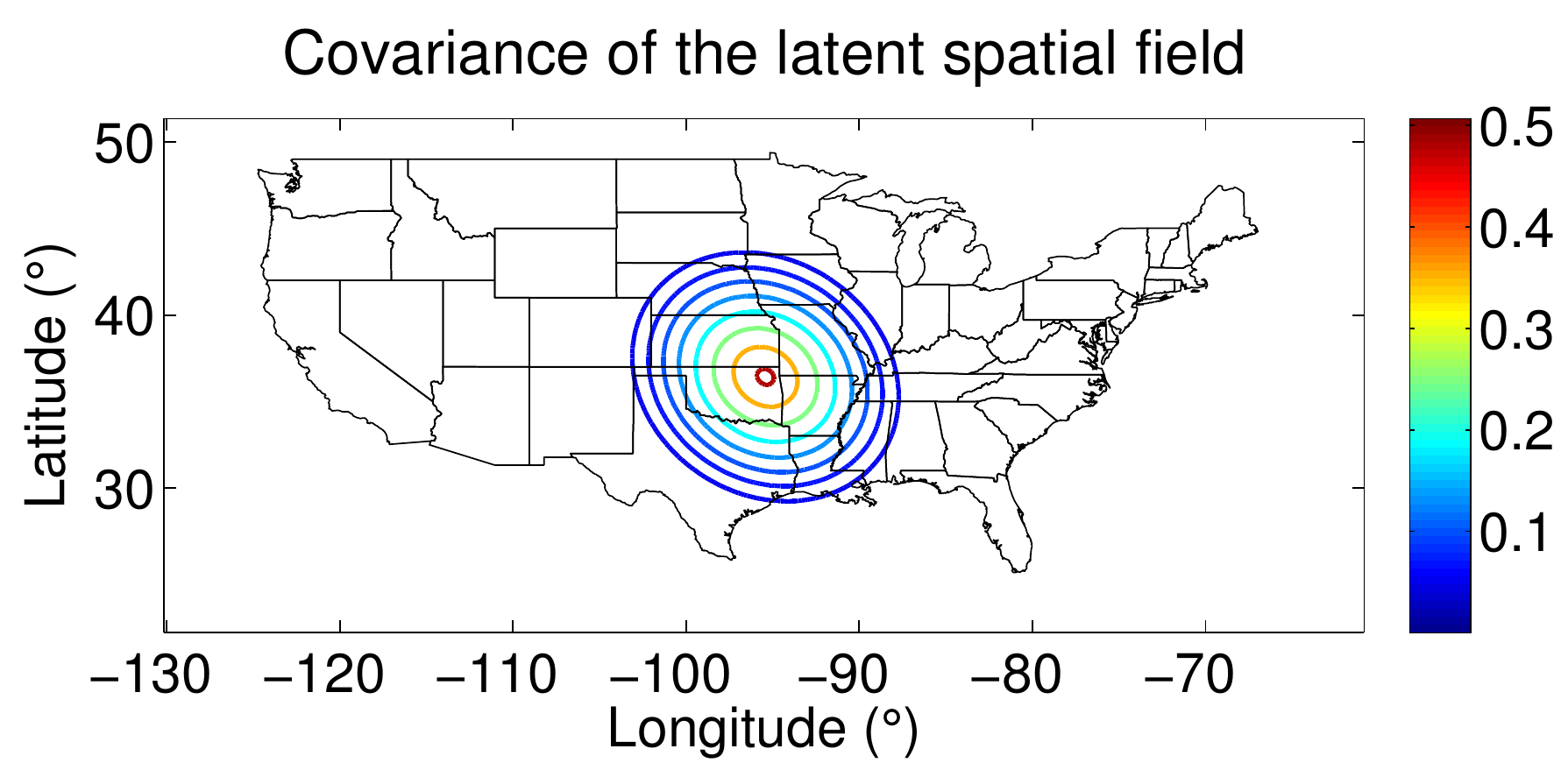}
	\caption{The 0.95, 0.70, 0.50, 0.36, 0.26, 0.19, 0.14 and 0.1 level correlation contours of the estimated covariance function for the stationary model.}
	\label{fig:statCov}
\end{figure}

From Section~\ref{sec:GRF} one can see that the estimated model implies a covariance
function approximately equal to the Mat\'{e}rn covariance function
\[
	r(\boldsymbol{s}_1,\boldsymbol{s}_2) = \hat{\sigma}^2 \left\lvert\left\lvert \left(\hat{\mathbf{H}}/\hat{\kappa}^2\right)^{-1/2}(\boldsymbol{s}_2-\boldsymbol{s}_1)\right\rvert\right\rvert K_1\left(\left\lvert\left\lvert \left(\hat{\mathbf{H}}/\hat{\kappa}^2\right)^{-1/2}(\boldsymbol{s}_2-\boldsymbol{s}_1)\right\rvert\right\rvert\right),
\]
where \(\hat{\sigma}^2 = 0.505\) and
\[
	\frac{\hat{\mathbf{H}}}{\hat{\kappa}^2} = \begin{bmatrix} 5.71 & -0.86 \\ -0.86 & 4.96 \end{bmatrix},
\] 
with an additional nugget effect with precision \(\hat{\tau}_\mathrm{noise} = 71.2\).
Figure~\ref{fig:statCov} shows contours of the estimated covariance function with 
respect to one location. One can see that the model gives high dependence within
a typical-sized state, whereas there is little dependence between the centres of
different typically-sized states.

\begin{figure}
	\centering
	\subfigure[Prediction]{
		\includegraphics[width=14cm]{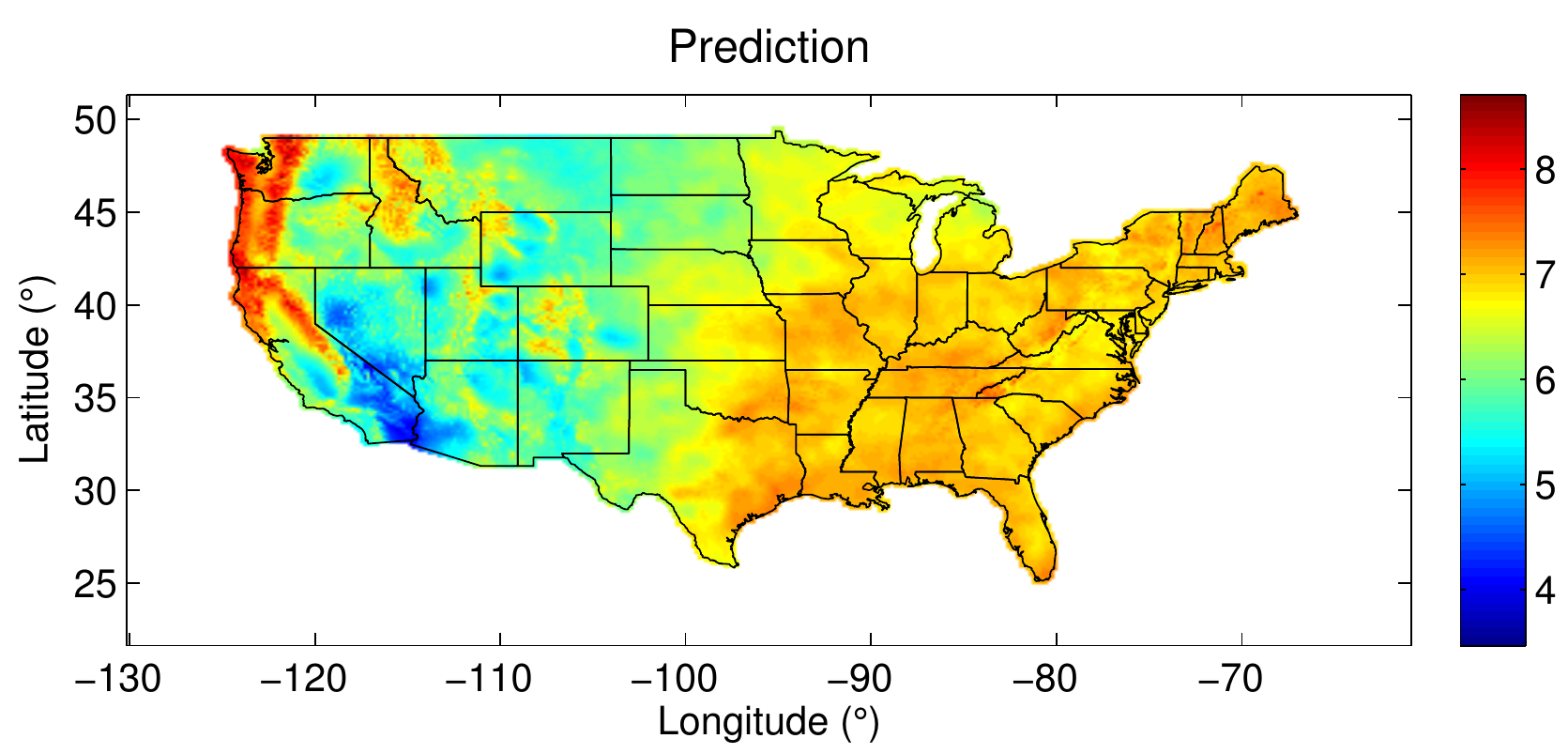}
	}\\
	\subfigure[Prediction standard deviations]{
		\includegraphics[width=14cm]{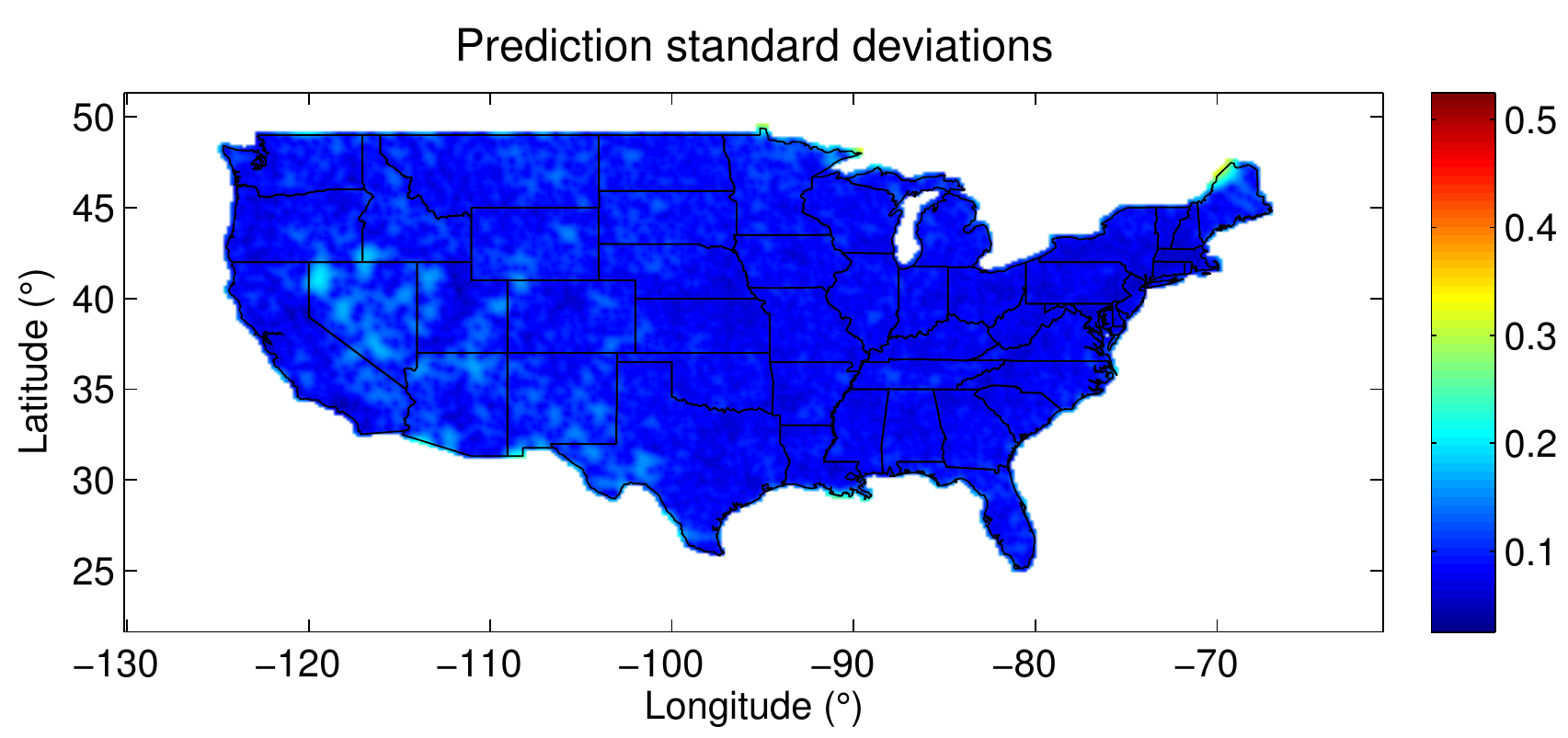}
		\label{fig:StatPred:StdDev}
	}
	\caption{Predicted values and prediction standard deviations for the
			 stationary model.}
	\label{fig:StatPred}
\end{figure}

Next, the parameter values are used together with the observed
logarithms of annual precipitations to predict the logarithm of
annual precipitation at the centre of each cell in the discretization.
This gives \(400\times 200\) locations to predict. The elevation covariate
for each location is selected from bilinear interpolation from the closest 
points in the high resolution elevation data set GLOBE~\cite{Globe1999}. 
The predictions and prediction standard deviations are shown in 
Figure~\ref{fig:StatPred}. Since the observations are contained within the
conterminous US, the locations outside are coloured white.

\subsection{Non-stationary model}
This section uses the same domain size and observations as in 
Section~\ref{sec:StatModel}.

\subsubsection{Selection of the smoothing parameters}
\label{sec:SmoothSel}
As discussed in Section~\ref{sec:fullModel}, the hyperparameters
\(\tau_1\), \(\tau_2\), \(\tau_3\) and \(\tau_4\), that appear in the
priors for the functions \(\log(\kappa^2)\), \(\log(\gamma)\), \(v_x\)
and \(v_y\), must be pre-selected in some way before the rest of the inference
is done. Since these hyperparameters control smoothing in the third stage of a
hierarchical spatial model, there is little information available
about them in the data. Attempts were made to give them 
Gamma priors and infer them 
together with \(\boldsymbol{\alpha}_1\),
\(\boldsymbol{\alpha}_2\), \(\boldsymbol{\alpha}_3\), \(\boldsymbol{\alpha}_4\)
and \(\log(\tau_{\mathrm{noise}})\) as MAP estimates, but this leads to 
estimates that are too influenced by the Gamma priors.

The hyperparameters are chosen with 5-fold cross-validation based on 
the log-predictive density. The data is randomly divided into five parts 
and in turn one part is used as test data and the other four parts are
used as training data. For each choice of \(\tau_1\), \(\tau_2\), \(\tau_3\) and
\(\tau_4\) the cross-validation error is calculated by
\[
	\mathrm{CV}(\tau_1, \tau_2, \tau_3, \tau_4) = -\frac{1}{5}\sum_{i=1}^{5} \log(\pi(\boldsymbol{y}_i^*|\boldsymbol{y}_i, {\skew{3}\hat{\boldsymbol{\theta}}}_i),
\]
where \(\boldsymbol{y}_i^*\) is the test data and \({\skew{3}\hat{\boldsymbol{\theta}}}_i\) is the MAP estimate of the parameters based
on the training data \(\boldsymbol{y}_i\) using the selected \(\tau\)-values. This function is
calculated for \(\log(\tau_i)\in\{2,4,6,8\}\) for 
\(i = 1, 2, 3, 4\). The model is run on a \(200 \times 100\) grid
with \(8 \times 4\) basis functions for each function. The 
choice that gave the smallest cross-validation error was
\(\log(\tau_1) = 2\), \(\log(\tau_2) = 4\), \(\log(\tau_3) = 2\) 
and \(\log(\tau_4) = 8\).

\subsubsection{Parameter estimates}
The non-stationary spatial model is constructed with the same discretization of the
domain as the stationary model. But in the non-stationary spatial model each of
the four functions in the SPDE is given a \(16 \times 8\) basis 
with a second-order random walk prior. The hyperparameters in the priors are
set to the values from Section~\ref{sec:SmoothSel}. Together with the precision
parameter of the random effect this gives a total of 513 parameters. These parameters
are estimated together with a MAP estimate. It is worth noting that there are not
513 ``free'' parameters as most of them are connected together in four different
Gaussian process priors. This means that an increase in the number of parameters
would mean increases in the resolutions of these Gaussian processes.

\begin{figure}
	\centering
	\subfigure[Approximate standard deviations]{
		\includegraphics[width=7cm]{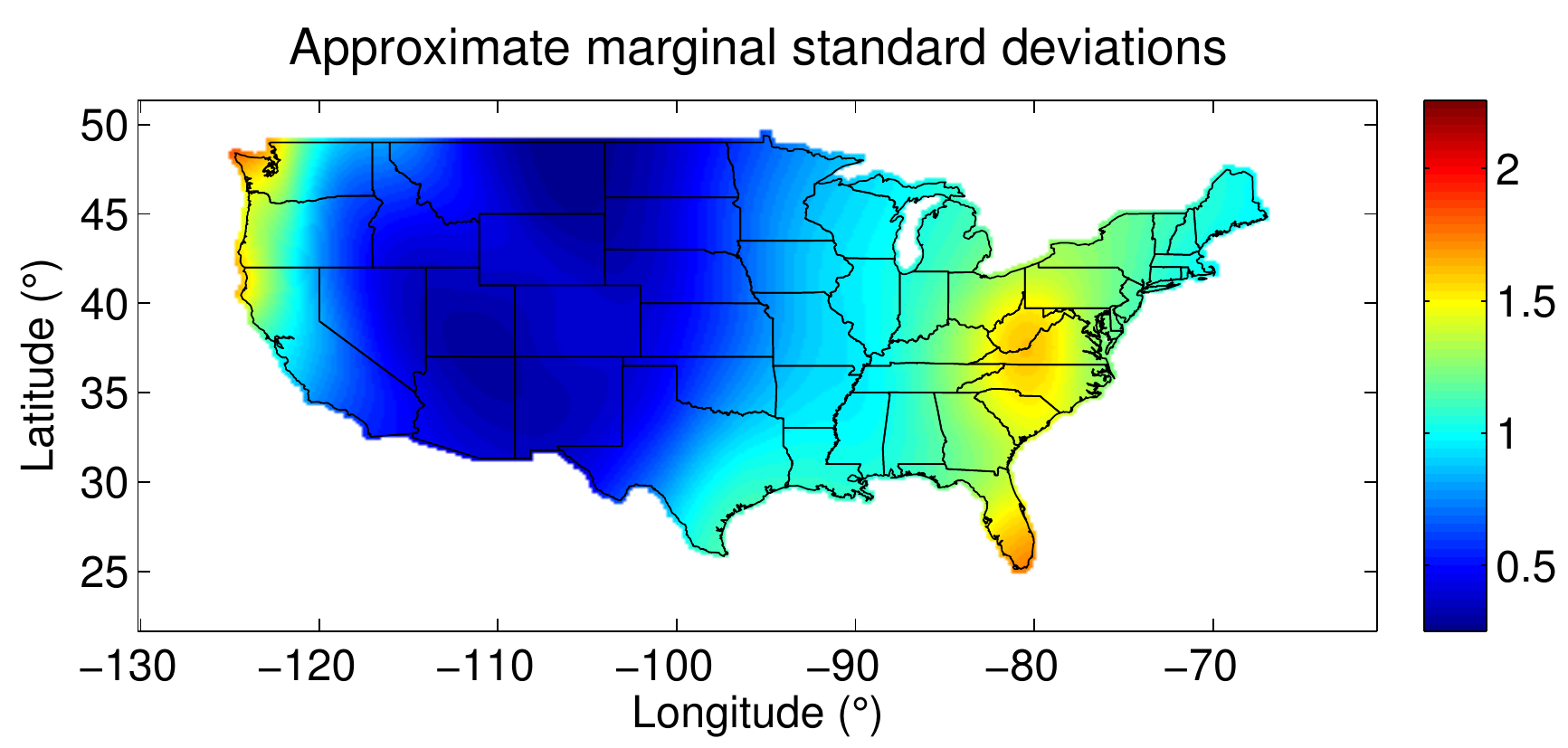}
		\label{fig:NonStatPar:StdDev}
	}
	\subfigure[Approximate correlation structure]{
		\includegraphics[width=7cm]{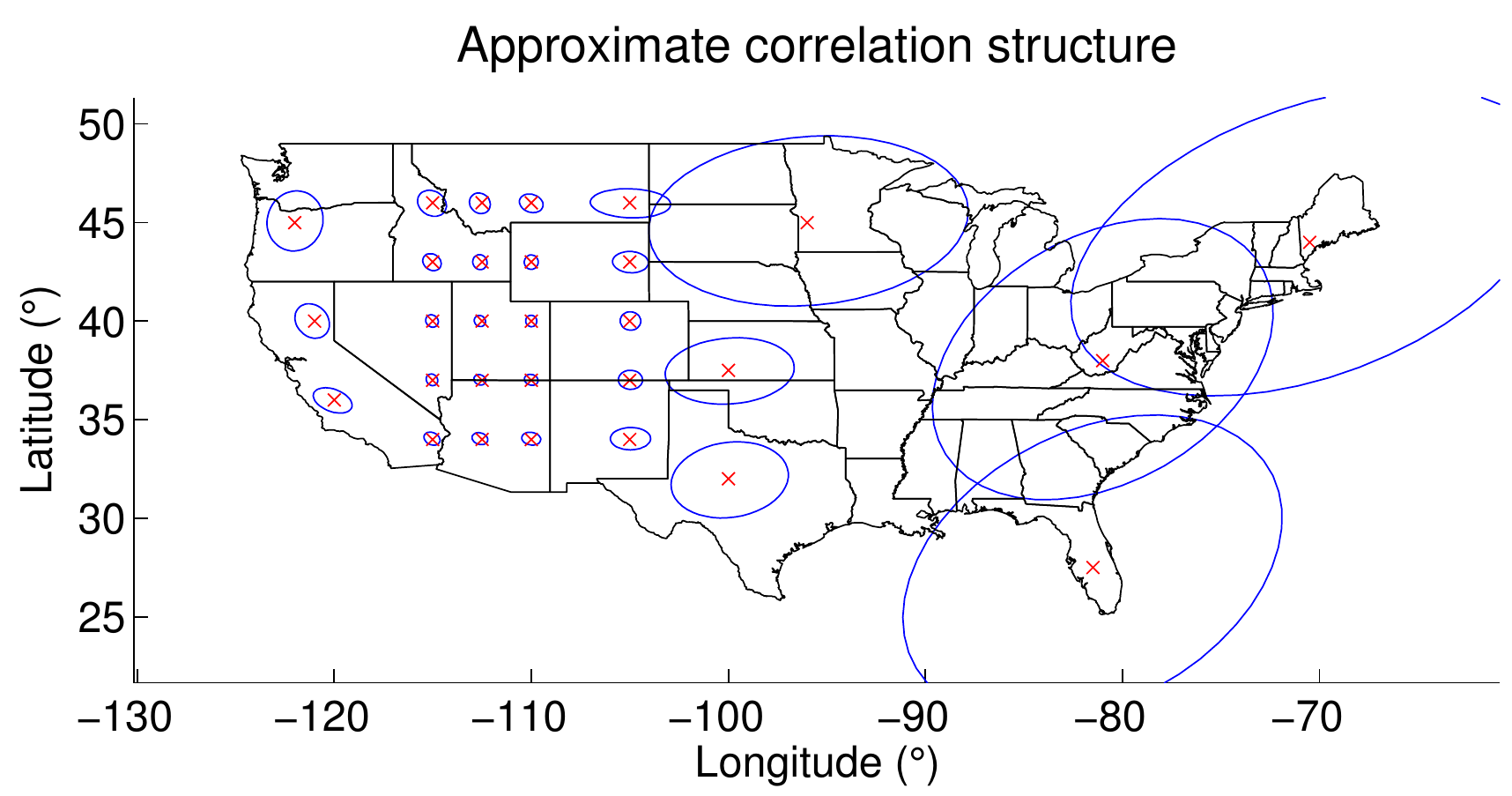}
		\label{fig:NonStatPar:Corr}
	}\\
	\subfigure[Exact marginal standard devitaions]{
		\includegraphics[width=7cm]{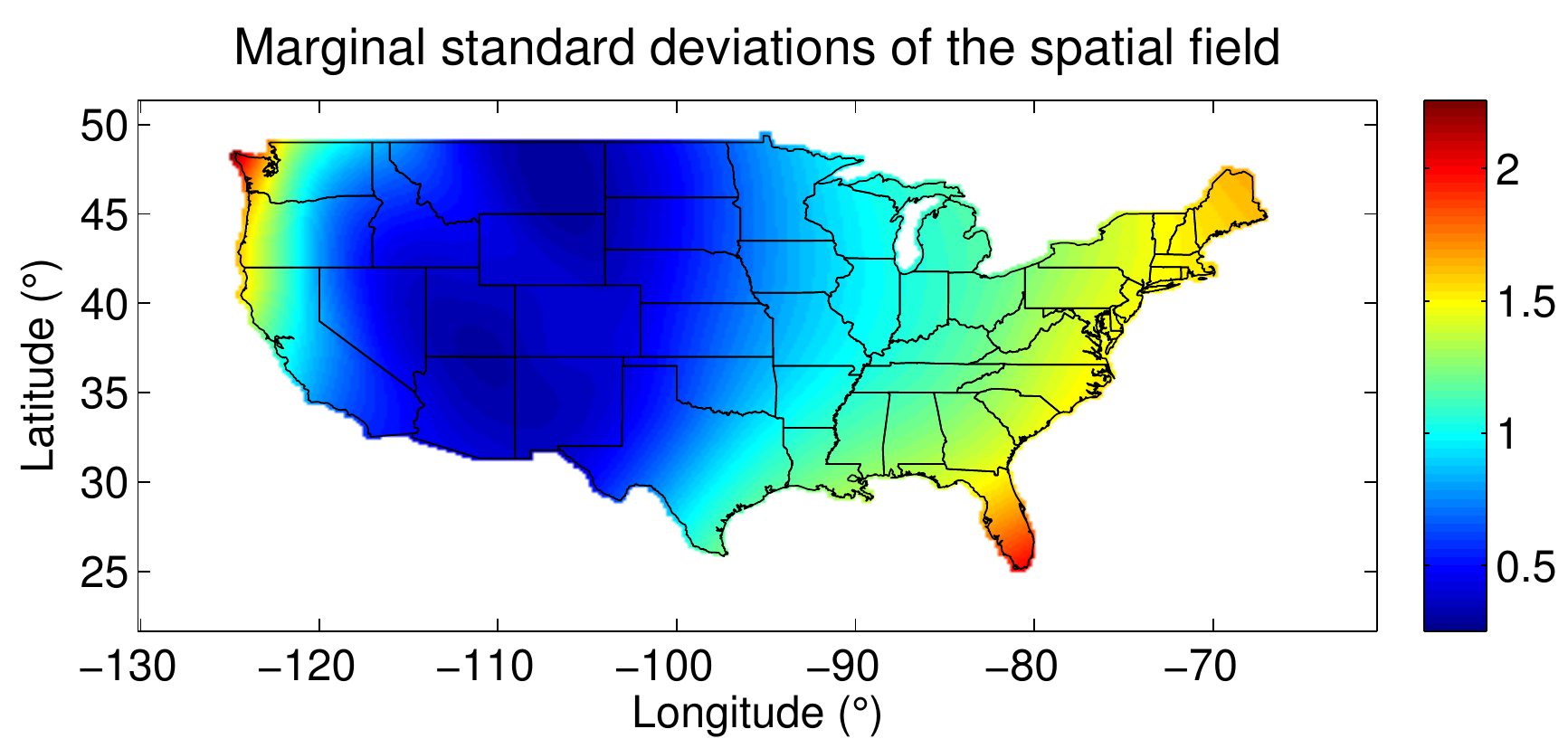}
		\label{fig:NonStatPar:TrueStdDev}
	}
	\subfigure[Exact correlation structure]{
		\includegraphics[width=7cm]{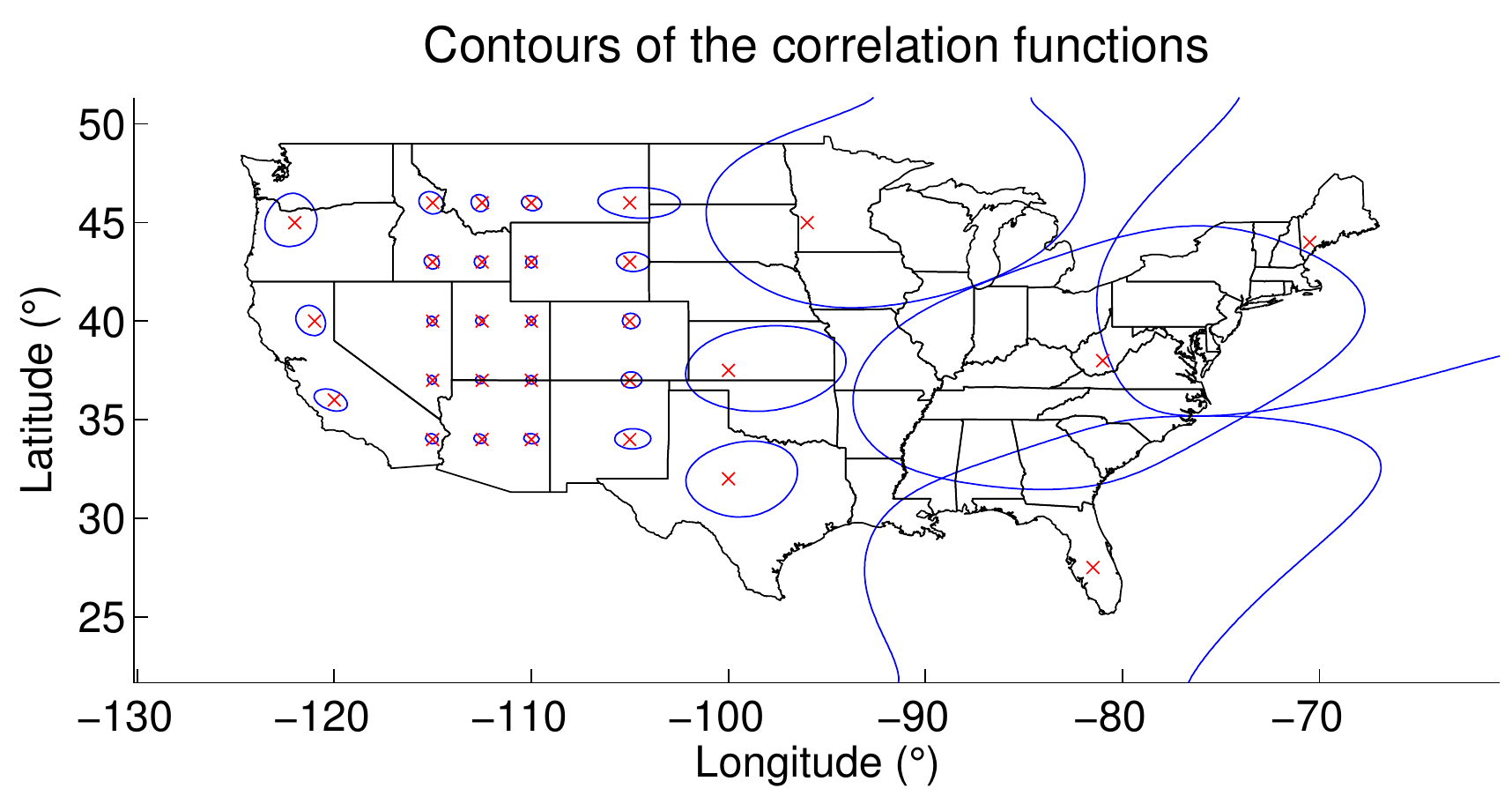}
		\label{fig:NonStatPar:TrueCorr}
	}
	\caption{Estimated covariance structure of the spatial field.
			 \subref{fig:NonStatPar:StdDev} Marginal standard deviations of the 
			 stationary models defined by the parameter values in each point
			 \subref{fig:NonStatPar:Corr} Contours of 0.7 correlation for the 
			 stationary models defined by the parameter values at selected locations
			 marked with red crosses
			 \subref{fig:NonStatPar:TrueStdDev} Exact marginal standard deviations
			 \subref{fig:NonStatPar:TrueCorr} Exact contours of 0.7 correlation
			 for selected locations marked with red crosses
			 }
	\label{fig:NonStatPar}
\end{figure}

The nugget effect has an estimated precision of 
\(\hat{\tau}_\mathrm{noise} = 107.4\), and the estimates 
of \(\kappa^2\) and \(\mathbf{H}\) are not shown since the
exact values themselves are not interesting. However, 
from the estimated functions it is possible to approximately
visualize the resulting covariance structure of the latent field. This
can be done with the covariance function in Equation~\eqref{eq:aniMatern} which 
describes the covariances of a stationary model. For each location in the 
grid the marginal standard deviation is calculated using the 
Mat{\'e}rn covariance function with the parameters at that location. 
This gives the results shown in Figure~\ref{fig:NonStatPar:StdDev}. Then for 
selected locations the correlation function defined by the parameters at 
those locations are used to draw 0.7 level correlation contours as shown 
in Figure~\ref{fig:NonStatPar:Corr}.

The figures based on stationary models give a quick indication of the
structure, but are only approximations. The exact marginal standard deviations 
and 0.7 level correlation contours for the same locations as above are given in 
Figure~\ref{fig:NonStatPar:TrueStdDev} and Figure~\ref{fig:NonStatPar:TrueCorr}.
From these figures one can see that there is
good correspondence between the approximations and the exact calculations.
It is interesting to observe that the exact correlation contours have the same
general shape, but are stretched corresponding to whether the range of the
stationary models are increasing or decreasing. The exact contours for 
the locations around longitude \(100\,^{\circ}\) are 
``larger'' in the east direction and ``smaller'' in the west direction than
the contours based on the stationary approximation.

It is worth mentioning that since only one realization is used, one cannot expect
the estimated covariance structure to be ``the true'' covariance structure. It is
impossible to separate the effects due to non-stationarity in the mean and 
the effects due to non-stationarity in the covariance structure. 
Therefore, the estimated structure must be understood to say something about both 
how well the covariates describe the data at different locations and
the non-stationarity in the covariance structure. In this case there is
a good fit for the elevation covariate in the mountainous areas in the western
part, but it offers less information in the eastern part. From 
Figure~\ref{fig:observation} one can see that at around longitude 
\(97^\circ\, \mathrm{W}\) there is an increase in precipitation 
which cannot be explained by elevation, and thus not captured by the
covariates. In the areas with long correlation ``range'', the 
spatial field is ``approaching'' a second-order random walk.

\begin{figure}
	\centering
	\subfigure[Marginal standard deviations]{
		\includegraphics[width=7cm]{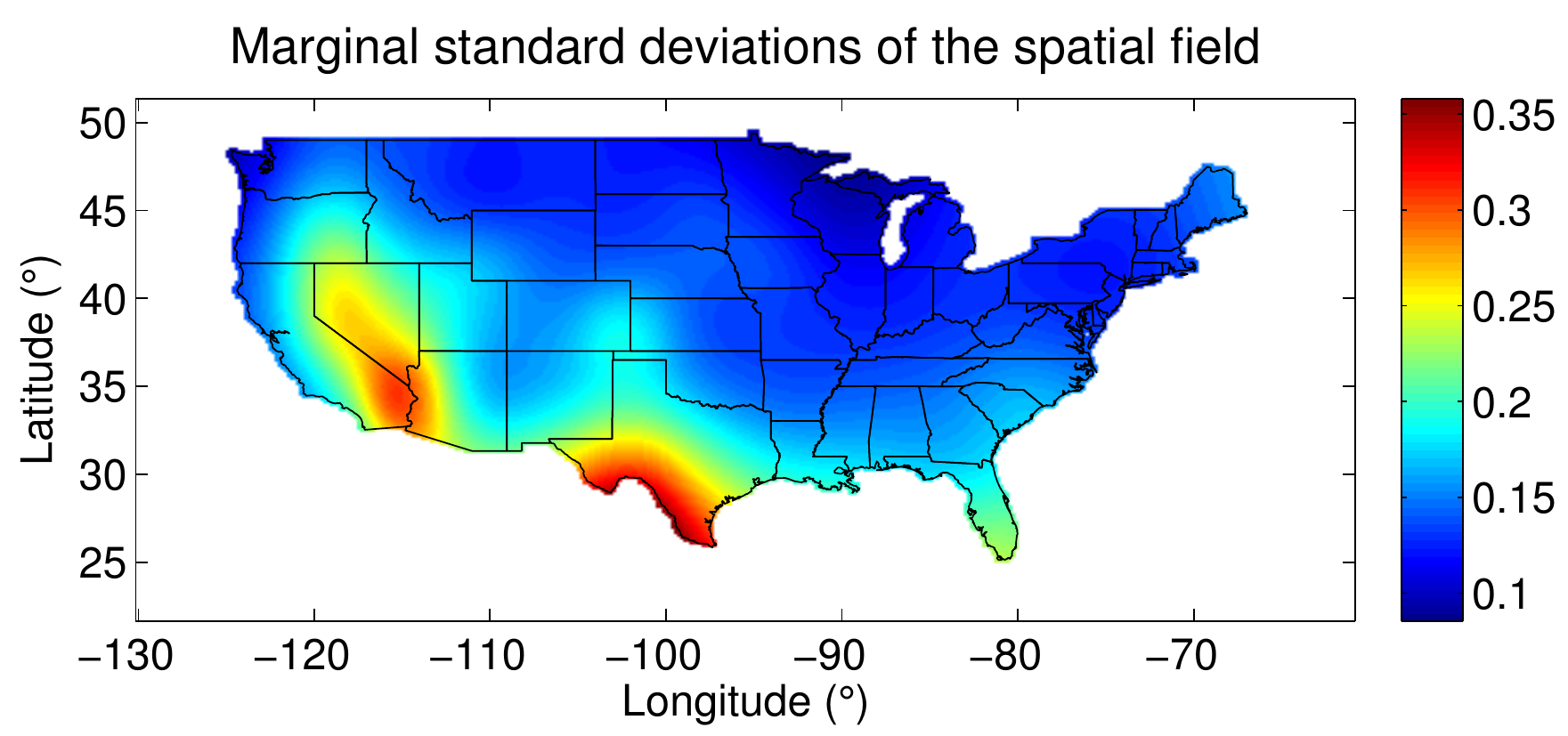}
		\label{fig:nonStatMu:StdDev}
	}
	\subfigure[Covariance structure]{
		\includegraphics[width=7cm]{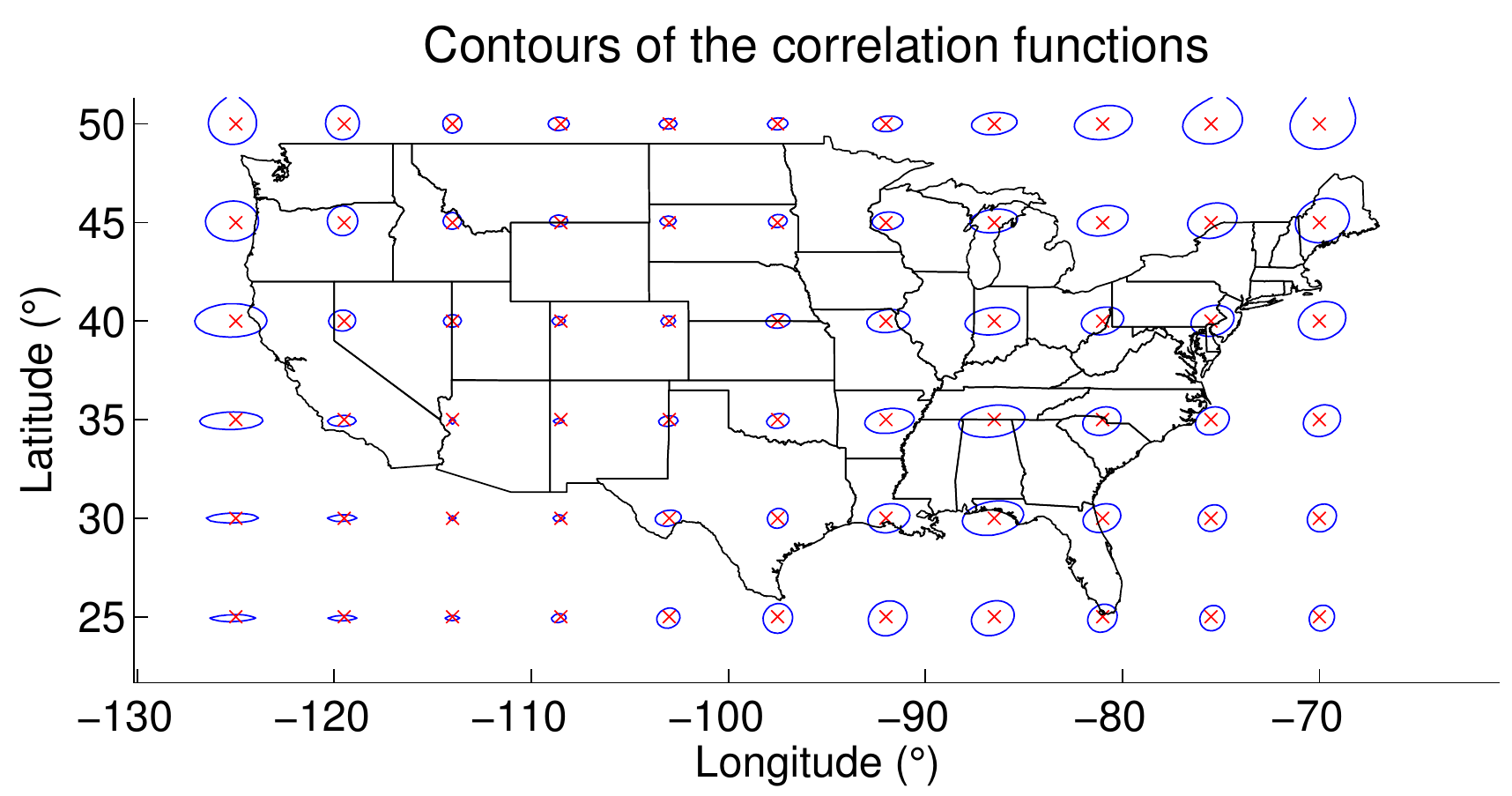}
		\label{fig:nonStatMu:Corr}
	}
	\caption{Estimate of the prior covariance structure with an extra covariate. 
			 \subref{fig:nonStatMu:StdDev} Marginal standard deviations
			 \subref{fig:nonStatMu:Corr} Contours of 0.7 level correlation for
			 selected locations marked with red crosses}
	\label{fig:nonStatMu}
\end{figure}

In the dataset used for this application there is more information 
available about the covariance structure than has been used. In addition
to 1981 there is precipitation data available for 102 other years. These
data could be used to create a full spatio-temporal model, but this is
not the intention of this paper. Instead the effect of the additional 
information will be illustrated by extracting a covariate which explains
much of the non-stationarity in the mean. For each location 
used in the 1981 dataset take the mean over the 102 other years 
(using the fill-in values as necessary).
This covariate used together with a joint mean gives the 
marginal standard deviations and correlation contours shown in 
Figure~\ref{fig:nonStatMu}.
One can see that the amount of non-stationarity captured by the mean has
a large impact on the resulting covariance structure, and one should
strive to include the information that is available. But one can see from
the Figure~\ref{fig:nonStatMu} that there is still evidence 
of non-stationarity in the covariance structure with this additional covariate.

\subsubsection{Prediction}
\begin{figure}
	\centering
	\subfigure[Prediction]{
		\includegraphics[width=14cm]{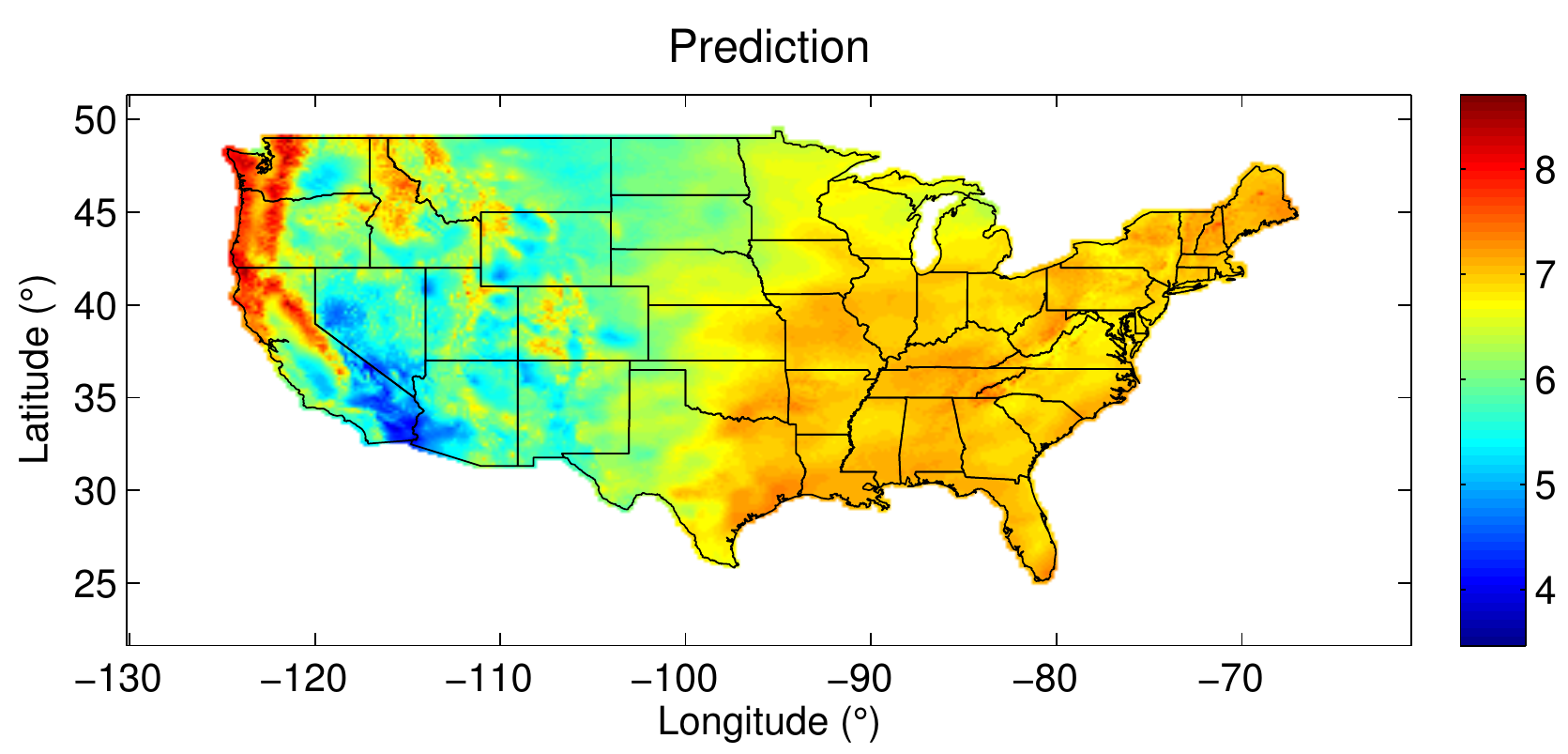}
		\label{fig:NonStatPred:Pred}
	}\\
	\subfigure[Prediction standard deviations]{
		\includegraphics[width=14cm]{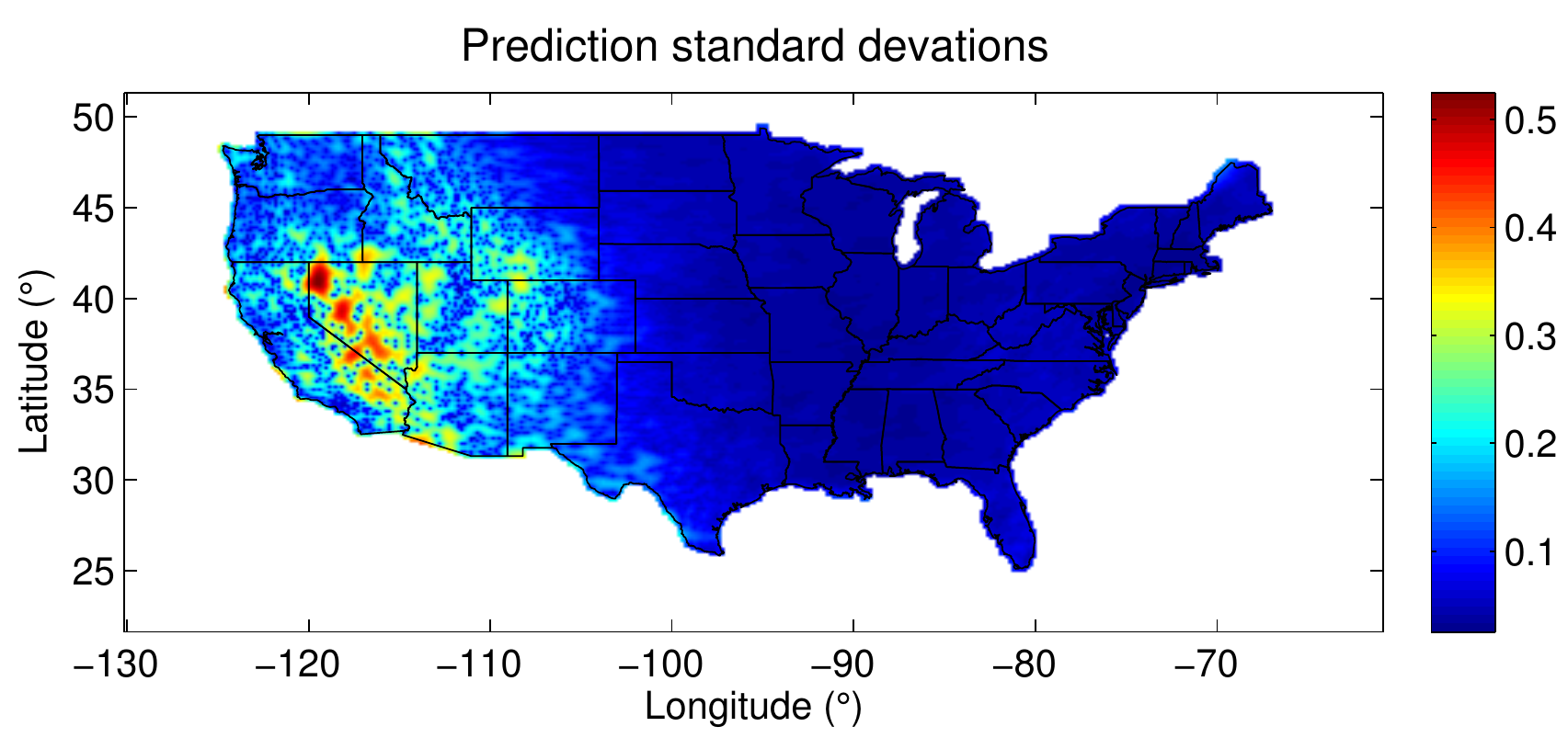}
		\label{fig:NonStatPred:StdDev}
	}
	\caption{Predictions and prediction standard deviations for the 
			 non-stationary model for the logarithm of annual precipitation
			 in the conterminous US in 1981 measured in millimetres.}
	\label{fig:NonStatPred}
\end{figure}

In the same way as in Section~\ref{sec:StatModel} the logarithm of annual
precipitation is predicted at the centre of each cell in the discretization. 
This gives predictions for \(400 \times 200\) regularly distributed locations, 
where the value of the elevation covariate at each location is selected with 
bilinear interpolation from the closest points in the GLOBE~\cite{Globe1999} dataset.
The prediction and prediction standard deviations are shown in 
Figure~\ref{fig:NonStatPred}. In the same way as for the 
stationary model, the values outside
the conterminous US are coloured white. One can see that the overall look of the 
predictions is similar to the predictions from the stationary model, 
but that the prediction standard deviations differ. In the non-stationary model
it varies how quickly the prediction standard deviations increase as one
moves away from an observed location.

\subsection{Comparison of the stationary and the non-stationary model}
The predictions of the stationary model and the non-stationary model are
compared with the continuous rank probability score (CRPS)~\cite{Gneiting2005} 
and the logarithmic
scoring rule. CRPS is defined for a univariate distribution as
\[
	\mathrm{crps}(F, y) = \int_{-\infty}^{\infty}\! (F(y)-\mathbbm{1}(y\leq t))^2\, \mathrm{d}t,
\]
where \(F\) is the distribution function of interest, \(y\) is an observation and
\(\mathbbm{1}\) is the indicator function. This gives a measure of how well
a single observation fits a distribution. The total score is calculated
as the average CRPS for the test data,
\[
	\mathrm{CRPS} = \frac{1}{N}\sum_{i=1}^{N}\mathrm{crps}(F_k, y_k),
\]
where \(\{y_k\}\) is the test data and \(\{F_k\}\) are the corresponding marginal prediction
distributions given the estimated covariance parameters and the training data. 
The logarithmic scoring rule is based on the joint distribution of the test data 
\(\boldsymbol{y}^*\) given the estimated covariance parameters 
\(\hat{\boldsymbol{\theta}}\) and the training data \(\boldsymbol{y}\),
\[
	\mathrm{LogScore} = -\log\pi\left(\boldsymbol{y}^*|\hat{\boldsymbol{\theta}},\boldsymbol{y}\right).
\]

\begin{figure}
	\centering
	\subfigure[Logarithmic score]{
		\includegraphics[height=5.5cm]{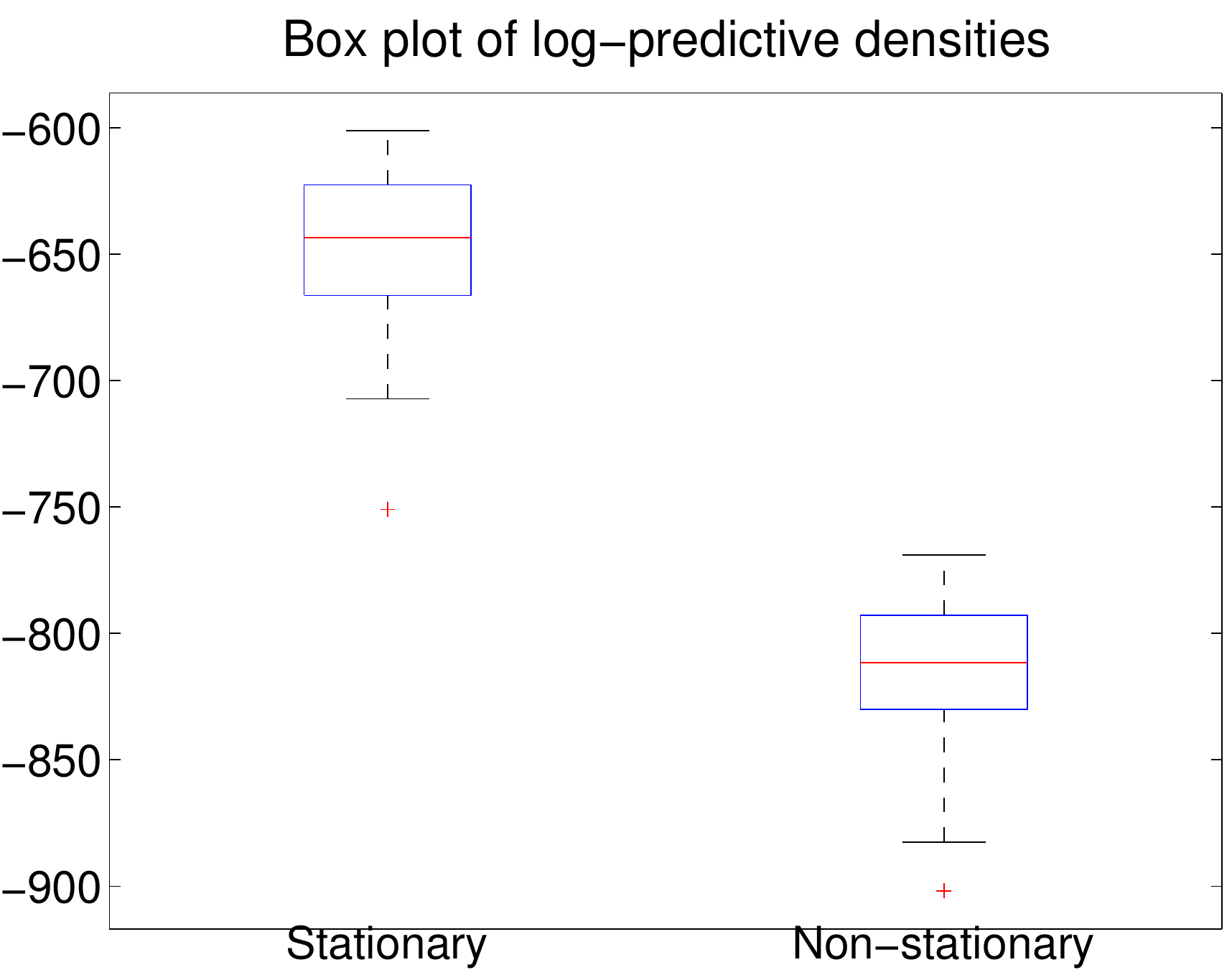}
		\label{fig:Comp:LPD}
	}
	\subfigure[CRPS]{
		\includegraphics[height=5.5cm]{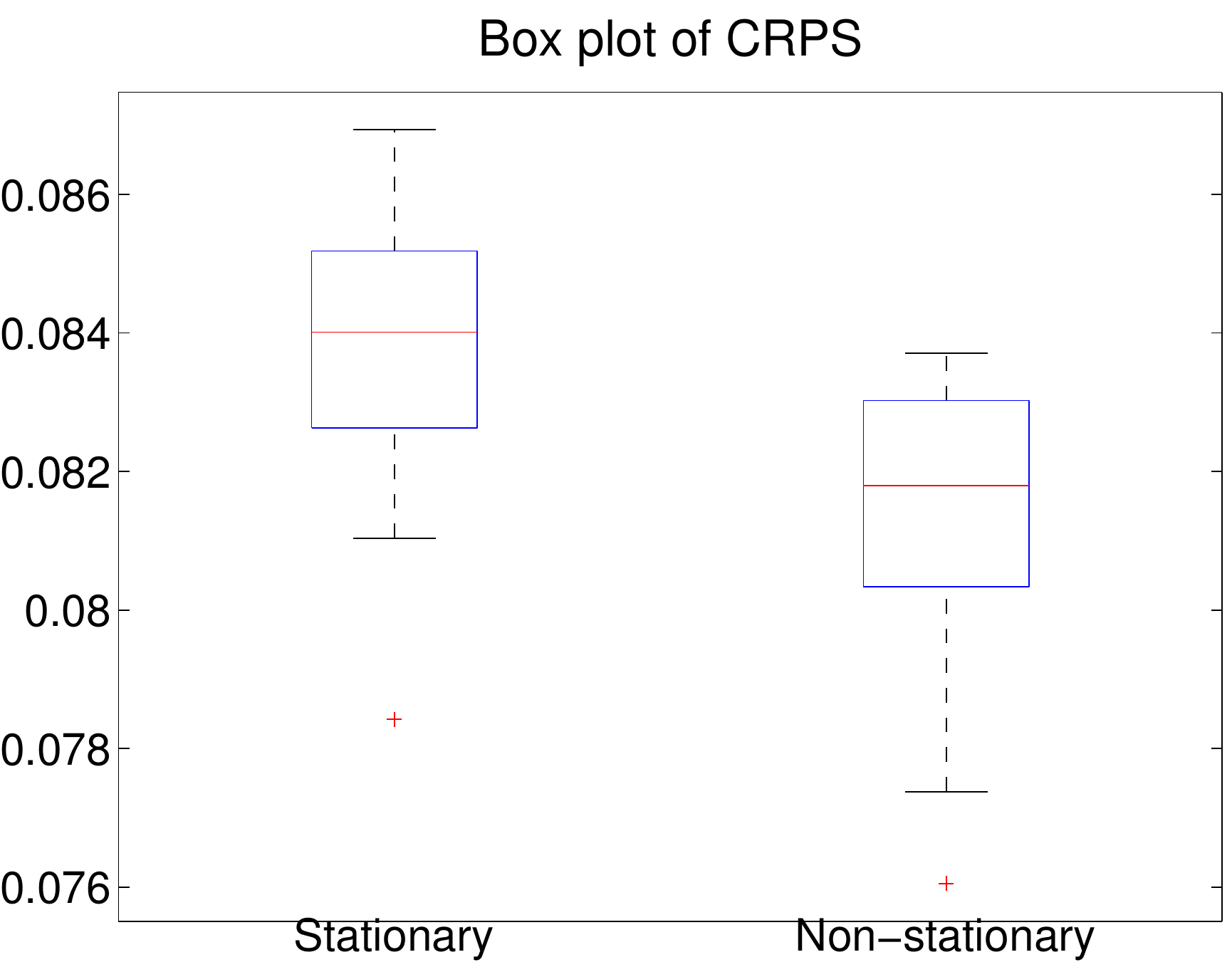}
		\label{fig:Comp:CRPS}
	}
	\caption{Box-plot of prediction scores from the stationary and
			 the non-stationary model. 20\% of the locations are randomly chosen
			 to be held out and the remaining 80\% are used to estimate parameters
			 and predict the 20\% held out data. This was repeated 20 times. Lower
			 is better.}
	\label{fig:Comp}
\end{figure}

The comparison is done with holdout sets. 
For each holdout set 20\% of the locations are chosen randomly. The remaining 
80\% of the locations are used to estimate the parameters and to predict the 
values at the locations in the holdout set. This procedure is repeated 20 times. 
For each repetition both the CRPS and the logarithmic score is calculated. 
From Figure~\ref{fig:Comp} one can see that measured by both scores 
the non-stationary model gives better predictions.

\section{Discussion and conclusion}
\label{sec:Discussion}
SPDE-based modelling offers a different point of view than
modelling based on covariance functions. The focus is shifted from global
properties to local properties. The object which is modelled
is the ``local'' dependences around a point, which are then automatically 
combined into a global structure. This offers some benefits in the sense
that it is easier to have intuition about local behaviour, but on the other
hand the SPDE introduces coefficients whose influence on the global structure
might not be immediately obvious. But it is possible to gain intuition about
the global behaviour through the stationary models defined by the parameters at each point.

In this paper it has been demonstrated that a non-stationary spatial model can
be constructed from an SPDE in such a way that the resulting model can be
estimated and used for prediction. The modelling can be done with an SPDE
while the computations are done with a GMRF approximation. This makes the
estimation computationally feasible for the number of observations used
in the application to annual precipitation. The model allows a flexible
structure which gives completely different covariance structures in the 
western and eastern part of the conterminous US. Additionally, the 
non-stationary model leads to better prediction than the stationary model 
measured both with CRPS and the logarithmic scoring rule.

The estimated covariance structure in itself is not that interesting since it
is a combination of unexplained non-stationarity in the mean and 
actual non-stationarity in the covariance structure. Non-stationarity 
from these two sources are indistinguishable with a single realization. 
If one includes a covariate in the mean that better explains the
non-stationarity, one ends up with a very different covariance structure. 
However, such extra information is not available in the spatial smoothing type 
application treated here.

The major challenge remaining is selecting the smoothing hyperparameters used 
in the priors for the coefficients of the SPDE. These control smoothing at
the third stage of a hierarchical spatial model and are not easily inferred.
Using cross-validation to select them is both an inefficient solution and an unsatisfactory
solution from a Bayesian point of view. The major problem is the number 
of parameters in the model, which makes it hard to study the marginal posterior
distribution of the smoothing parameters. It might be of interest to look into
simulation based methods.

\newpage
\appendix
\numberwithin{equation}{section}
\section{Derivation of the second-order random walk prior}
\label{app:Prior}
Each function, \(f\), is a priori modelled as a Gaussian process described by the 
SPDE
\begin{equation}
	-\Delta f(\boldsymbol{s}) = \frac{1}{\sqrt{\tau}}\mathcal{W}(\boldsymbol{s}), \qquad \boldsymbol{s}\in \mathcal{D} = [A_1, B_1]\times[A_2, B_2], 
	\label{Prior:eq:SPDE}
\end{equation}
where \(A_1 < B_1\), \(A_2 < B_2\) and \(\tau>0\), \(\mathcal{W}\) is standard
Gaussian white noise and 
\(\Delta = \frac{\partial^2}{\partial x^2}+\frac{\partial^2}{\partial y^2}\), with
the Neumann boundary condition of zero normal derivatives at the edges. In practice
this is approximated by representing \(f\) as a linear combination of basis elements
\(\{f_{ij}\}\) weighted by Gaussian distributed weights 
\(\{\alpha_{ij}\}\),
\[
	f(\boldsymbol{s}) = \sum_{i=1}^K \sum_{j=1}^L \alpha_{ij} f_{ij}(\boldsymbol{s}).
\]
The basis functions are constructed from separate bases \(\{g_i\}\) and \(\{h_j\}\)
for the \(x\)-coordinate and the \(y\)-coordinate, respectively,
\begin{equation}
	f_{ij}(\boldsymbol{s}) = g_i(x) h_j(y).
	\label{Prior:eq:fBase}
\end{equation}
For convenience each basis function is assumed to fulfil the boundary condition of
zero normal derivative at the edges.

Let \(\boldsymbol{\alpha} = \mathrm{vec}([\alpha_{ij}]_{ij})\), then the task is to
find the best Gaussian distribution for \(\boldsymbol{\alpha}\). Where ``best'' is
used in the sense of making the resulting distribution for \(f\) ``close'' to a solution of SPDE~\eqref{Prior:eq:SPDE}. This is done by a least-squares approach 
where the vector created from doing inner products of the left hand side with 
\(-\Delta f_{kl}\)  must be equal in distribution to the vector created from doing 
the same to the right hand side,
\begin{equation}
	\mathrm{vec}\left([\langle -\Delta f, -\Delta f_{kl} \rangle_\mathcal{D}]_{kl}\right) \eqd \mathrm{vec}\left([\langle \mathcal{W}, -\Delta f_{kl} \rangle_\mathcal{D}]_{kl}\right).
	\label{Prior:eq:alpha}
\end{equation}

First, calculate the inner product that is needed
\begin{align*}
	\left\langle -\Delta g_ih_j, -\Delta g_k h_l\right\rangle_\mathcal{D} &= \left\langle \Delta g_i h_j, \Delta g_i h_j\right\rangle_\mathcal{D}\\
	&= \left\langle \left(\frac{\partial^2}{\partial x^2}g_i\right) h_j + g_i \frac{\partial^2}{\partial y^2} h_j, \left(\frac{\partial^2}{\partial x^2}g_{k}\right)h_l + g_k \frac{\partial^2}{\partial y^2} h_{l} \right\rangle_\mathcal{D}.
\end{align*}
The bilinearity of the inner product can be used to expand the expression in a
sum of four innerproducts. Each of these inner products can then be written as a product of two inner products. Due to lack of space this is not done explicitly, but one of these terms is, for example,
\[
	\left\langle\left(\frac{\partial^2}{\partial x^2}g_i\right) h_j, \left(\frac{\partial^2}{\partial x^2}g_k\right)h_l\right\rangle_\mathcal{D} = \left\langle\frac{\partial^2}{\partial x^2}g_i, \frac{\partial^2}{\partial y^2}g_k\right\rangle_{[A_1,B_1]}\left\langle h_j, h_l\right\rangle_{[A_2,B_2]}.
\] 
By inserting Equation~\eqref{Prior:eq:fBase} into Equation~\eqref{Prior:eq:alpha}
and using the above derivations together with integration by parts one can see that
the left hand side becomes
\[
	\mathrm{vec}\left([\langle -\Delta f, -\Delta f_{kl} \rangle_\mathcal{D}]_{kl}\right) = \mathbf{C}\boldsymbol{\alpha},
\]
where \(\mathbf{C} = \mathbf{G}_{2}\otimes \mathbf{H}_{0}+2\mathbf{G}_{1}\otimes \mathbf{H}_{1}+\mathbf{G}_{0}\otimes\mathbf{H}_{2}\) with
\[
	\mathbf{G}_{n} = \left[\left\langle \frac{\partial^n}{\partial x^n} g_i, \frac{\partial^n}{\partial x^n} g_j \right\rangle_{[A_1,B_1]}\right]_{i,j}
\]
and
\[
	\mathbf{H}_{n} = \left[\left\langle \frac{\partial^n}{\partial y^n} h_i, \frac{\partial^n}{\partial y^n} h_j \right\rangle_{[A_2, B_2]}\right]_{i,j}	.
\]

The right hand side is a Gaussian random vector where the covariance between the
position corresponding to \(\alpha_{ij}\) and the position corresponding to 
\(\alpha_{kl}\) is given by
\[
	\langle -\Delta f_{ij}, -\Delta f_{kl} \rangle_\mathcal{D}.
\]
Thus the covariance matrix of the right hand side must be \(\mathbf{C}\)
and Equation~\eqref{Prior:eq:alpha} can be written in matrix form as
\[
	\mathbf{C}\boldsymbol{\alpha} = \mathbf{C}^{1/2}\boldsymbol{z},	
\]
where \(\boldsymbol{z}\sim\mathcal{N}_{KL}(\boldsymbol{0}, \mathbf{I}_{KL})\).
This means that \(\boldsymbol{\alpha}\) should be given the precision matrix
\(\mathbf{Q} = \mathbf{C}\). Note that \(\mathbf{C}\) might be singular
due to invariance to some linear combination of the basis elements.

\section{Conditional distributions}
\label{app:CondDist}
From the hierarchical model
\begin{align*}
	\text{Stage 1: } & \boldsymbol{y}|\boldsymbol{z}, \boldsymbol{\theta} \sim \mathcal{N}_N(\mathbf{S}\boldsymbol{z}, \mathbf{I}_N/\tau_{\mathrm{noise}})  \\
	\text{Stage 2: } & \boldsymbol{z}|\boldsymbol{\theta} \sim \mathcal{N}_{mn+p}(\boldsymbol{0}, \mathbf{Q}_z^{-1}),
\end{align*}
the posterior distribution \(\pi(\boldsymbol{\theta}|\boldsymbol{y})\) can be derived
explicitly. There are three steps involved.

\subsection{Step 1}
Calculate the distribution \(\pi(\boldsymbol{z}|\boldsymbol{\theta}, \boldsymbol{y})\)
up to a constant,
\begin{align*}
	\pi(\boldsymbol{z}|\boldsymbol{\theta}, \boldsymbol{y}) & \propto \pi(\boldsymbol{z}, \boldsymbol{\theta}, \boldsymbol{y}) \\
		& = \pi(\boldsymbol{\theta}) \pi(\boldsymbol{z}|\boldsymbol{\theta}) \pi(\boldsymbol{y}|\boldsymbol{\theta}, \boldsymbol{z}) \\
		& \propto \exp\left(-\frac{1}{2}(\boldsymbol{z}-\boldsymbol{0})^\mathrm{T}\mathbf{Q}_z(\boldsymbol{z}-\boldsymbol{0})-\frac{1}{2}(\boldsymbol{y}-\mathbf{S}\boldsymbol{z})^\mathrm{T}\mathbf{I}_N\cdot\tau_{\mathrm{noise}}(\boldsymbol{y}-\mathbf{S}\boldsymbol{z})\right)\\
		& \propto \exp\left(-\frac{1}{2}\left(\boldsymbol{z}^\mathrm{T}(\mathbf{Q}_z+\tau_{\mathrm{noise}}\mathbf{S}^\mathrm{T}\mathbf{S})\boldsymbol{z} - 2\boldsymbol{z}^\mathrm{T}\mathbf{S}^\mathrm{T}\boldsymbol{y}\cdot\tau_{\mathrm{noise}}\right)\right) \\
		& \propto \exp\left(-\frac{1}{2}(\boldsymbol{z}-\boldsymbol{\mu}_\mathrm{C})^\mathrm{T}\mathbf{Q}_\mathrm{C}(\boldsymbol{z}-\boldsymbol{\mu}_\mathrm{C})\right),
\end{align*}
where \(\mathbf{Q}_\mathrm{C} = \mathbf{Q}_z+\mathbf{S}^\mathrm{T}\mathbf{S}\cdot\tau_{\mathrm{noise}}\) and \(\mu_\mathrm{C} = \mathbf{Q}_\mathrm{C}^{-1}\mathbf{S}^\mathrm{T}\boldsymbol{y}\cdot \tau_{\mathrm{noise}}\). This is recognized as a Gaussian distribution
\[
	\boldsymbol{z}| \boldsymbol{\theta}, \boldsymbol{y} \sim \mathcal{N}_N(\boldsymbol{\mu}_\mathrm{C}, \mathbf{Q}_\mathrm{C}^{-1}).
\]

\subsection{Step 2}
\label{app:CondDist:2}
Integrate out \(\boldsymbol{z}\) from the joint distribution of \(\boldsymbol{z}\),
\(\boldsymbol{\theta}\) and \(\boldsymbol{y}\) via the Bayesian rule,

\begin{align*}
	\pi(\boldsymbol{\theta},\boldsymbol{y}) &= \frac{\pi(\boldsymbol{\theta}, \boldsymbol{z}, \boldsymbol{y})}{\pi(\boldsymbol{z}|\boldsymbol{\theta}, \boldsymbol{y})} \\
		& = \frac{\pi(\boldsymbol{\theta})\pi(\boldsymbol{z}|\boldsymbol{\theta})\pi(\boldsymbol{y}| \boldsymbol{z}, \boldsymbol{\theta})}{\pi(\boldsymbol{z}|\boldsymbol{\theta}, \boldsymbol{y})}.
\end{align*}
The left hand side of the expression does not depend on the value of
\(\boldsymbol{z}\), therefore the right hand side may be evaluated
at any desired value of \(\boldsymbol{z}\). Evaluating at 
\(\boldsymbol{z} = \boldsymbol{\mu}_\mathrm{C}\) gives
\begin{align*}
	\pi(\boldsymbol{\theta}, \boldsymbol{y}) & \propto \frac{\pi(\boldsymbol{\theta})\pi(\boldsymbol{z} = \boldsymbol{\mu}_\mathrm{C})\pi(\boldsymbol{y}|\boldsymbol{z} = \boldsymbol{\mu}_\mathrm{C}, \boldsymbol{\theta})}{\pi(\boldsymbol{z}=\boldsymbol{\mu}_\mathrm{C}|\boldsymbol{\theta}, \boldsymbol{y})} \\
		& \propto\pi(\boldsymbol{\theta}) \frac{|\mathbf{Q}_z|^{1/2}|\mathbf{I}_N\cdot \tau_{\mathrm{noise}}|^{1/2}}{|\mathbf{Q}_\mathrm{C}|^{1/2}}\exp\left(-\frac{1}{2}\boldsymbol{\mu}_\mathrm{C}^\mathrm{T}\mathbf{Q}_z\boldsymbol{\mu}_\mathrm{C}\right)\times\\
		& \times \exp\left(-\frac{1}{2}(\boldsymbol{y}-\mathbf{S}\boldsymbol{\mu}_\mathrm{C})^\mathrm{T}\mathbf{I}_N\cdot\tau_{\mathrm{noise}}(\boldsymbol{y}-\mathbf{S}\boldsymbol{\mu}_\mathrm{C})\right)\times\\
		& \times \exp\left(+\frac{1}{2}(\boldsymbol{\mu}_\mathrm{C}-\boldsymbol{\mu}_\mathrm{C})^\mathrm{T}\mathbf{Q}_\mathrm{C}(\boldsymbol{\mu}_\mathrm{C}-\boldsymbol{\mu}_\mathrm{C})\right).
\end{align*}

\subsection{Step 3}
Condition on \(\boldsymbol{y}\) to get the desired conditional distribution,
\begin{align}
	\notag \log(\pi(\boldsymbol{\theta}|\boldsymbol{y})) &= \mathrm{Const} + \log(\pi(\boldsymbol{\theta}))+\frac{1}{2}\log(\det(\mathbf{Q}_z))+\frac{N}{2}\log(\tau_\mathrm{noise})+ \\
	& -\frac{1}{2}\log(\det(\mathbf{Q}_\mathrm{C}))-\frac{1}{2}\boldsymbol{\mu}_\mathrm{C}^\mathrm{T}\mathbf{Q}_z\boldsymbol{\mu}_z-\frac{\tau_\mathrm{noise}}{2}(\boldsymbol{y}-\mathbf{S}\boldsymbol{\mu}_\mathrm{C})^\mathrm{T}(\boldsymbol{y}-\mathbf{S}\boldsymbol{\mu}_\mathrm{C}).
	\label{eq:CondDist:post}
\end{align}

\section{Analytic expression for the gradient}
\label{app:Gradient}
This appendix shows the derivation of the derivative of the log-likelihood.
Choose the evaluation point \(\boldsymbol{z} = \boldsymbol{0}\)
in Appendix~\ref{app:CondDist:2} to find
\begin{align*}
	\log(\pi(\boldsymbol{\theta},\tau_\mathrm{noise}|\boldsymbol{y})) &= \mathrm{Const} + \log(\pi(\boldsymbol{\theta},\tau_\mathrm{noise}))+\frac{1}{2}\log(\det(\mathbf{Q}_z))+\frac{N}{2}\log(\tau_\mathrm{noise})+ \\
	& -\frac{1}{2}\log(\det(\mathbf{Q}_\mathrm{C}))-\frac{\tau_\mathrm{noise}}{2}\boldsymbol{y}^\mathrm{T}\boldsymbol{y}+\frac{1}{2}\boldsymbol{\mu}_\mathrm{C}^\mathrm{T}\mathbf{Q}_\mathrm{C}\boldsymbol{\mu}_\mathrm{C}.
\end{align*}
This is just a rewritten form of Equation~\eqref{eq:CondDist:post} which is more 
convenient for the calculation of the gradient, and which separates the 
\(\tau_\mathrm{noise}\) parameter from the rest of the covariance parameters. First
some preliminary results are presented, then the derivatives are calculated with
respect to \(\theta_i\) and lastly the derivatives are calculated with respect to
\(\log(\tau_\mathrm{noise})\).

Begin with simple preliminary formulas for the derivatives of the conditional 
precision matrix with respect to each of the parameters,
\begin{equation}
	\frac{\partial}{\partial \theta_i} \mathbf{Q}_\mathrm{C} = \frac{\partial}{\partial \theta_i}(\mathbf{Q}+\mathbf{S}^\mathrm{T}\mathbf{S}\cdot\tau_\mathrm{noise}) = \frac{\partial}{\partial \theta_i}\mathbf{Q}
	\label{eq:Gradient:Qc_theta}
\end{equation}
and
\begin{equation}
	\frac{\partial}{\partial \log(\tau_{\mathrm{noise}})} \mathbf{Q}_\mathrm{C} = \frac{\partial}{\partial \log(\tau_{\mathrm{noise}})}(\mathbf{Q}+\mathbf{S}^\mathrm{T}\mathbf{S}\cdot\tau_\mathrm{noise}) = \mathbf{S}^\mathrm{T}\mathbf{S}\cdot\tau_\mathrm{noise}.
	\label{eq:Gradient:Qc_tau}
\end{equation}

\subsection{Derivative with respect to \(\theta_i\)}
First the derivatives of the log-determinants can be handled by an explicit
formula~\cite{petersen2012} 
\begin{align*}
	\frac{\partial}{\partial \theta_i} (\log(\det(\mathbf{Q}))-\log(\det(\mathbf{Q}_\mathrm{C})) &= \mathrm{Tr}(\mathbf{Q}^{-1}\frac{\partial}{\partial \theta_i}\mathbf{Q})-\mathrm{Tr}(\mathbf{Q}_\mathrm{C}^{-1}\frac{\partial}{\partial \theta_i} \mathbf{Q}_\mathrm{C})\\
	&= \mathrm{Tr}\left[(\mathbf{Q}^{-1}-\mathbf{Q}_\mathrm{C}^{-1})\frac{\partial}{\partial \theta_i}\mathbf{Q}\right].
\end{align*}
Then the derivative of the quadratic forms are calculated
\begin{align*}
	\frac{\partial}{\partial \theta_i} \left(-\frac{1}{2}\boldsymbol{y}^\mathrm{T}\boldsymbol{y}\cdot\tau_\mathrm{noise} + \frac{1}{2}\boldsymbol{\mu}_\mathrm{C}\mathbf{Q}_\mathrm{C}\boldsymbol{\mu}_\mathrm{C} \right) &= 0 + \frac{\partial}{\partial \theta_i} \left(\frac{1}{2}\boldsymbol{y}^\mathrm{T}\tau_\mathrm{noise}\mathbf{S}\mathbf{Q}_\mathrm{C}^{-1}\mathbf{S}^\mathrm{T}\tau_\mathrm{noise}\boldsymbol{y}\right)\\
	&= -\frac{1}{2}\boldsymbol{y}^\mathrm{T}\tau_\mathrm{noise}\mathbf{S}\mathbf{Q}_\mathrm{C}^{-1}\left(\frac{\partial}{\partial \theta_i}\mathbf{Q}_\mathrm{C}\right)\mathbf{Q}_\mathrm{C}^{-1}\mathbf{S}^\mathrm{T}\tau_\mathrm{noise}\boldsymbol{y}  \\
	&= -\frac{1}{2}\boldsymbol{\mu}_\mathrm{C}^\mathrm{T}\left(\frac{\partial}{\partial \theta_i}\mathbf{Q}\right)\boldsymbol{\mu}_\mathrm{C}.
\end{align*}
Combining these gives
\[
	\frac{\partial}{\partial \theta_i} \log(\pi(\boldsymbol{\theta}, \tau_\mathrm{noise}| \boldsymbol{y})) = \frac{\partial}{\partial \theta_i} \log(\pi(\boldsymbol{\theta},\tau_\mathrm{noise} )) + \frac{1}{2}\mathrm{Tr}\left[(\mathbf{Q}^{-1}-\mathbf{Q}_\mathrm{C}^{-1})\frac{\partial}{\partial \theta_i} \mathbf{Q}\right] - \frac{1}{2}\boldsymbol{\mu}_\mathrm{C}^\mathrm{T}\left(\frac{\partial}{\partial \theta_i}\mathbf{Q}\right)\boldsymbol{\mu}_\mathrm{C}
\]

\subsection{Derivative with respect to \(\log(\tau_\mathrm{noise})\)}
First calculate the derivative of the log-determinants
\begin{align*}
	\frac{\partial}{\partial \log(\tau_\mathrm{noise})} \left(N\log(\tau_\mathrm{noise}) - \log(\det(\mathbf{Q}_\mathrm{C}))\right) &= N - \mathrm{Tr}\left(\mathbf{Q}_\mathrm{C}^{-1}\frac{\partial}{\partial \log(\tau_\mathrm{noise})} \mathbf{Q}_\mathrm{C}\right)\\
		& = N-\mathrm{Tr}\left(\mathbf{Q}_\mathrm{C}^{-1}\mathbf{S}^\mathrm{T}\mathbf{S}\cdot\tau_\mathrm{noise}\right).
\end{align*}
Then the derivative of the quadratic forms
\begin{align*}
	\frac{\partial\left(-\frac{1}{2}\boldsymbol{y}^\mathrm{T}\boldsymbol{y}\cdot\tau_\mathrm{noise} + \frac{1}{2} \boldsymbol{\mu}_\mathrm{C}\mathbf{Q}_\mathrm{C}\boldsymbol{\mu}_\mathrm{C}\right)}{\partial \log(\tau_\mathrm{noise})} &= -\frac{1}{2}\boldsymbol{y}^\mathrm{T}\boldsymbol{y}\cdot\tau_\mathrm{noise} + \frac{\partial}{\partial \log(\tau_n)} \frac{1}{2}\boldsymbol{y}^\mathrm{T}\tau_\mathrm{noise}\mathbf{S}\mathbf{Q}_\mathrm{C}^{-1}\mathbf{S}^\mathrm{T}\tau_\mathrm{noise}\boldsymbol{y}\\
	&= -\frac{1}{2}\boldsymbol{y}^\mathrm{T}\boldsymbol{y}\cdot\tau_\mathrm{noise} +\boldsymbol{y}^\mathrm{T}\tau_\mathrm{noise}\mathbf{S}\mathbf{Q}_\mathrm{C}^{-1}\mathbf{S}\left(\frac{\partial \tau_{\mathrm{noise}}}{\partial \log(\tau_{\mathrm{noise}})}\right)\boldsymbol{y}+ \\
	&- \frac{1}{2}\boldsymbol{y}^\mathrm{T}\tau_\mathrm{noise}\mathbf{S}\mathbf{Q}_\mathrm{C}^{-1}\left(\frac{\partial}{\partial \log(\tau_{\mathrm{noise}})} \mathbf{Q}_\mathrm{C} \right)\mathbf{Q}_\mathrm{C}^{-1}\mathbf{S}^\mathrm{T}\tau_\mathrm{noise}\boldsymbol{y}\\
	& = -\frac{1}{2}\boldsymbol{y}^\mathrm{T}\boldsymbol{y}\cdot\tau_\mathrm{noise} +\boldsymbol{\mu}_\mathrm{C}^\mathrm{T}\mathbf{S}^\mathrm{T}\boldsymbol{y}\cdot\tau_\mathrm{noise} - \frac{1}{2}\boldsymbol{\mu}_\mathrm{C}^\mathrm{T}\mathbf{S}^\mathrm{T} \mathbf{S}\boldsymbol{\mu}_\mathrm{C}\cdot\tau_\mathrm{noise}\\
	&= -\frac{1}{2}(\boldsymbol{y}-\mathbf{A}\boldsymbol{\mu}_\mathrm{C})^\mathrm{T}(\boldsymbol{y}-\mathbf{A}\boldsymbol{\mu}_\mathrm{C})\cdot\tau_\mathrm{noise}.
\end{align*}
Together these expressions give
\begin{align*}
	\frac{\partial \log(\pi(\boldsymbol{\theta}, \tau_\mathrm{noise}| \boldsymbol{y}))}{\partial \log(\tau_\mathrm{noise})}  &= \frac{\partial}{\partial \log(\tau_\mathrm{noise})} \log(\pi(\boldsymbol{\theta},\tau_\mathrm{noise})) + \frac{N}{2} - \frac{1}{2}\mathrm{Tr}\left[\mathbf{Q}_\mathrm{C}^{-1}\mathbf{S}^\mathrm{T}\mathbf{S}\cdot\tau_\mathrm{noise}\right]+ \\
	&-\frac{1}{2}(\boldsymbol{y}-\mathbf{A}\boldsymbol{\mu}_\mathrm{C})^\mathrm{T}(\boldsymbol{y}-\mathbf{A}\boldsymbol{\mu}_\mathrm{C})\cdot\tau_\mathrm{noise}
\end{align*}

\subsection{Implementation}
The derivative $\frac{\partial}{\partial \theta_i} \mathbf{Q}_c$ can be 
calculated quickly since it is a simple functions of $\theta$. The trace of 
the inverse of a matrix \(A\) times the derivative of a matrix \(B\) only requires the
values of the inverse of \(A\) for non-zero elements of \(B\). In the above case
the two matrices have the same type of non-zero structure, but it can happen that
specific elements in the non-zero structure are zero for one of the matrices.
This way of calculating the inverse only at a subset of the locations can be handled as described
in~\ocite{Gelfand2010}*{Sections~12.1.7.10--12.1.7.12}.

\bibliography{references}

\end{document}